\documentclass[aps,superscriptaddress,nofootinbib,showpacs,eqsecnum,preprint,tightenlines]{revtex4}
\usepackage{hyperref}
\usepackage{epsfig,rotating}
\usepackage{amsmath,amssymb}
\usepackage{dsfont}
\usepackage{bbm}
\usepackage{slashed}
\numberwithin{equation}{section}

\newcommand{\be}{\begin{equation}}
\newcommand{\ee}{\end{equation}}
\newcommand{\bea}{\begin{eqnarray}}
\newcommand{\eea}{\end{eqnarray}}

\newcommand{\vx}{\vec{x}}

\newcommand{\vp}{\vec{p}}

\newcommand{\vq}{\vec{q}}

\newcommand{\vk}{\vec{k}}

\begin{document}

\title{Nearly degenerate heavy sterile neutrinos in cascade decay: mixing and oscillations.}

\author{Daniel Boyanovsky}
\email{boyan@pitt.edu}
 \affiliation{Department of Physics and
Astronomy, University of Pittsburgh, Pittsburgh, PA 15260}

\date{\today}

\begin{abstract}

 Some extensions beyond the Standard Model propose the existence of nearly degenerate heavy sterile neutrinos. If kinematically allowed  these can be resonantly produced and decay in a cascade  to common final states. The common decay channels   lead to   mixing of the heavy sterile neutrino states  and interference effects.  We implement non-perturbative methods to study the dynamics of the  cascade decay to common final states, which features similarities but also noteworthy differences with the case of neutral meson mixing. We show that mixing and oscillations among the nearly degenerate sterile neutrinos can be detected as \emph{quantum beats} in the distribution  of final states  produced from their decay. These oscillations would be a telltale signal of mixing between   heavy sterile neutrinos. We study in detail the case of two   nearly degenerate sterile neutrinos produced in the decay of pseudoscalar mesons and decaying into a purely leptonic ``visible'' channel: $\nu_h \rightarrow e^+ e^- \nu_a$. Possible cosmological implications for the effective number of neutrinos $N_{eff}$ are discussed.

\end{abstract}


\maketitle

\section{Introduction}

Many extensions of the Standard Model that propose explanations of neutrino masses via see-saw type mechanisms\cite{saw1,saw2,saw3,saw4}   predict the existence of heavy ``sterile'' neutrinos namely $SU(2)\times U(1)$ singlets that mix very weakly with ``active'' neutrinos \cite{book1,revbilenky,book2,book3,grimuslec,mohapatra,degouvea,bilenky,book5}.  Heavy sterile neutrinos may play an important role in baryogenesis through leptogenesis\cite{yanagida,pilares,pila1,pila2} or via neutrino oscillations\cite{akhruba} motivating several models for leptogenesis which may also yield dark matter candidates\cite{asa1,revs}.  Furthermore, heavy sterile neutrinos may contribute to the energy transport during SNII explosions\cite{fullkuseSN}, their decay may be a source of early reionization\cite{earlyre}, they have been argued to play an  important role in the thermal history of the early Universe and to contribute to the cosmological neutrino background\cite{fullkuse}.  For a review of the role of sterile neutrinos in   cosmology and astrophysics  see refs.\cite{kuserev,revs,gorbunov,dolgovcosmo}.

   If the mass of the heavy sterile neutrino $m_h \lesssim M_{\pi,K},M_\tau$ they can be produced as resonances in the decay of pseudoscalar mesons  (or charged leptons)  opening a   window for current and future experimental searches.   A comprehensive study of leptonic and semileptonic weak decays of heavy sterile-like neutrinos was carried out in ref.\cite{shrock} and extended in ref.\cite{rosner}, and various experimental studies searching for heavy neutral leptons\cite{exp1,exp2,exp3,exp4,exp5,exp6,exp7,exp8,exp9,exp10,exp11,exp12,exp13,exp14,exp15} provide constraints on the values of the mixing matrix elements between heavy sterile and active neutrinos for a wide range of masses with   stringent bounds within the mass range $140\,\mathrm{MeV}\leq M_h \leq 500\,\mathrm{MeV}$\cite{exp12}.    Recent bounds on the mixing matrix elements between active (light) and sterile (heavy) neutrinos\cite{exp12,kuseexpt2,gronau} yield $|U_{eh}|^2;|U_{\mu h}|^2 \lesssim 10^{-7}-10^{-5}$ in the mass range $30 \,\mathrm{MeV} \lesssim m_h \lesssim 300 \, \mathrm{MeV}$.  If heavy sterile neutrinos are Majorana, they can  mediate lepton number violating transitions with $|\Delta l| =2$ motivating further studies of their production and decay\cite{tao,dib,castro}.  Furthermore, resonant production and mixing of \emph{nearly degenerate} heavy sterile neutrinos may lead to enhanced CP violation and baryogenesis\cite{pilares,pila1,pila2,akhruba,asa1,revs}.
    A thorough analysis of production and decay rates and cross sections of heavy neutral leptons in various mass regimes is available in refs.\cite{shrock,rosner,tao,dib,propo1,propo2,propo3,propo4,lou,masjuan}, providing the theoretical backbone to current and proposed experimental searches.

\vspace{2mm}

\textbf{Motivation and goals:}

The astrophysical, cosmological and phenomenological importance of heavy sterile neutrinos and their ubiquitous place in well motivated extensions beyond the Standard Model motivates   a series of recent proposals\cite{propo1,propo2,propo3,propo4,lou,masjuan}. These make a compelling case for rekindling the search for heavy sterile neutrinos in various current and next generation experiments.

 As pointed out in refs.\cite{pilares,pila1,pila2,akhruba,asa1,revs} extensions beyond the Standard Model that feature nearly degenerate heavy sterile neutrinos provide mechanisms for resonantly enhanced CP-violation with important consequences for baryogenesis through leptogenesis.
  If these nearly degenerate heavy sterile neutrinos are produced resonantly they may decay in a cascade into common channels leading to mixing\cite{pilares,pila1,pila2,segre}.
 Mixing and the ensuing time   dependent oscillation phenomena associated with the decay of (nearly) degenerate states into a common channel is a hallmark of the dynamics of neutral meson mixing such as $K^0 \overline{K^0}, B^0\overline{B^0}$\cite{kabir,bigi,lavoura}.

 The goal of this article is to explore in detail the mixing of two heavy but nearly degenerate sterile neutrinos   as a consequence of a common decay channel, the  concomitant \emph{time dependent oscillations} from their interference and the  observational consequences in the distribution of the decay products.

  Previous discussions of particle mixing focused either on the self-energy corrections featuring off diagonal matrix elements because of common intermediate states\cite{pilares,pila1,pila2} or effective Hamiltonian descriptions akin to the case of neutral meson mixing\cite{segre,kabir,bigi,lavoura}.

  Our goal is complementary in that we study the complete time evolution from the decay of an initial unstable state into   channels that include the nearly degenerate  heavy sterile neutrinos, which in turn  decay  into the final states, and assess the impact of the interference between the nearly degenerate states upon the distribution of final states.

 For this purpose we implement a systematic  quantum field theoretical generalization of the Wigner-Weisskopf approach\cite{cascade,mio} that includes the decay dynamics of the initial state and the time evolution of the final states. We consider the case of \emph{two} nearly degenerate heavy sterile neutrinos produced   from the decay of a pseudoscalar meson (or a heavy charged lepton) first within a general framework of cascade decay to common final states, and then consider the explicit case of  a purely leptonic ``visible'' decay channel for the heavy sterile neutrinos as a potential observable in future experiments.

 We find that while there are similarities with the case of neutral meson mixing ($K^0\overline{K^0};B^0\overline{B^0})$, there are important differences primarily as a consequence of the production of the heavy steriles from the decay of a parent particle (here a pseudoscalar meson) and also from the decay of the nearly degenerate heavy neutrinos into the final states.

\section{General formulation.}\label{sec:form}

We generalize the framework described in refs.\cite{cascade,mio} to describe the production, evolution and decay of two   heavy sterile neutrinos.

 Consider a  total Hamiltonian   $H=H_0+H_I$ with $H_0$ the free field Hamiltonian and
\be H_I = H_{\mathcal{P}} +H_{\mathcal{D}}+H_{ct}\label{Hint}\ee where $H_{\mathcal{P}}\,,\,H_{\mathcal{D}}$ refer generically to the production $(\mathcal{P})$ and decay $(\mathcal{D})$ interaction vertices and $H_{ct}$ refers to local renormalization counterterms.


To be specific, and motivated by current and future neutrino experiments,  we consider the case where sterile neutrinos are produced in the decay of a charged pseudoscalar meson $\Phi=\pi,K$ into a charged lepton $\alpha$ and a neutrino $i$ where   $i=a$, refers to the ``active-like'' (light) and $i=h$ to the ``sterile-like'' heavy neutrinos \emph{mass eigenstates}, with

\be
H_{\mathcal{P}} =i\,\frac{F_{\Phi}}{2} \sum_{\alpha=e,\mu} \sum_{i}  U_{\alpha i }\int d^3x \left[  \,\overline{\Psi}_{l_\alpha} (\vx,t)\, \gamma^{\mu}(1-\gamma^5)\Psi_{ {\nu_i}} (\vx,t)  \,    \partial_{\mu} \Phi (\vx,t) \right] + h.c.\,, \label{Hintfi}
\ee with
\be F_{\pi} = \sqrt{2}\, G_F\, V_{ud} \, f_{\pi} ~~;~~ F_{K} =  \sqrt{2}\, G_F V_{us}\, f_{K} \label{Fs} \ee where $f_{\pi};f_{K}$ are the corresponding decay constants and $ U_{\alpha i }$ is the neutrino mixing matrix with $i=a,h$.

 Specifically, the decay interaction vertex $H_\mathcal{D} $ is taken to be the usual Standard Model charged current and neutral current vertices, namely   $H_\mathcal{D}=H_{CC}+H_{NC}$  written in  the neutrino mass basis.


 Although we consider these specific production and decay vertices for the main discussion in this article, the formulation is more general and applicable for any other production and decay interaction Hamiltonians beyond the Standard Model. To make the discussion general, we consider the   case in which $H_\mathcal{D}$ describes the decay of $\nu_h$ into a multiparticle final state $\{  X \}$ ($\nu_h \rightarrow \{ X\}$).

 Let us consider an initial state   with one $\Phi$ meson of momentum $\vec{k}$ and the vacuum for the other fields, namely (to simplify  we use the same notation for the spatial Fourier transform of a field)

\be \big|\Psi(t=0)\rangle = \big|\Phi_{\vec{k}} \rangle \,. \label{inistate} \ee

\begin{figure}[h!]
\begin{center}
\includegraphics[height=4.5in,width=4.5in,keepaspectratio=true]{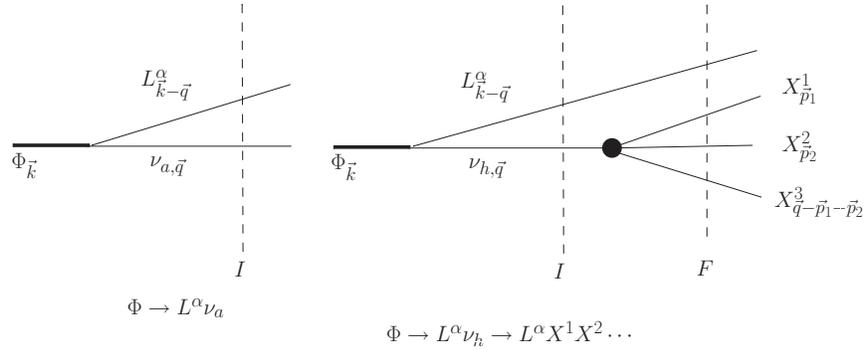}
\caption{Decay $\Phi \rightarrow L^\alpha\,\nu_a$ (left) and cascade decay $\Phi \rightarrow L^\alpha\, \nu_h \rightarrow L^\alpha\,\big\{X\big\}$ (right) where $\big\{X\big\}= X^1_{\vec{p}_1} X^2_{\vec{p}_2} X^3_{\vec{p}_3}\cdots$ is a multiparticle state with $\vec{p}_1+\vec{p}_2+\vec{p}_3+\cdots = \vec{q}$. The dashed lines depict the intermediate two particle state (I) and the final   multi  particle state (F).  }
\label{fig:cascade}
\end{center}
\end{figure}

Upon time evolution in the Schroedinger picture this state evolves into $\big|\Psi(t)\rangle$ obeying
\be \frac{d}{dt} \big|\Psi(t)\rangle_S = -i (H_0+H_I) \big|\Psi(t)\rangle_S\,. \label{timeevol} \ee
When $M_\Phi > m_{L^\alpha}+m_{\nu_h}~;~m_{\nu_h} >  m_{X}$ where $m_{X}$ is the invariant mass of the multiparticle final state $\{X\}$ the interaction Hamiltonian (\ref{Hint}) describes the cascade process depicted in fig.\ref{fig:cascade}.

We now pass to the interaction picture wherein
\be   H_I(t)= e^{iH_0t}\,H_I\,e^{-iH_0t} \label{HIint}\ee
and the state obeys
\be \label{inttimeevol}
i \frac{d}{dt} |\Psi(\vk,t)\rangle_I = H_I(t)|\Psi(\vk,t)\rangle_I \,.
\ee
Consider that at $t=0$ the initial state is the single meson state of spatial momentum $\vk$ given by (\ref{inistate}),  at any later time, the  state $|\Psi(\vk,t)\rangle_I$ is   expanded in the basis   $|n\rangle$ of  eigenstates of $H_0$, namely
\be |\Psi(\vk,t)\rangle_I = \sum_n A_n(t) |n\rangle \,. \label{ip}\ee

Up to second order in the interaction, the cascade decay depicted in fig. (\ref{fig:cascade}) is described by the following multiparticle state

\bea \label{intstate}
|\Psi(\vk,t)\rangle_I & = &  A_\Phi(\vec{k},t)\big|\Phi_{\vec{k}}\rangle + \sum_{\alpha;\vq;i=a,h}\,A^{\alpha\, i}_{I}(\vk,\vq;t)\,\big|\nu_{i,\vq};\,L^\alpha_{\vk-\vq} \rangle \nonumber \\ & + &    \sum_{\alpha;\vq;\{X\};\{\vec{p} \}_X} A^{\alpha\,X}_{F}(\vk,\vq,\{\vp\}_X;t)\,\big|L^\alpha_{\vk-\vq}\,; \{X\} \rangle  +\cdots
\eea

For simplicity of notation we do not distinguish between neutrino and antineutrino, furthermore,  the framework discussed below is general, independent of whether neutrinos are Dirac or Majorana.

In the last term in (\ref{intstate}), the sum over $\{X\}$ is over all the decay channels of $\nu_h$ and for each channel the sum over $\{\vp\}_X$ is over the momenta $\vp_1;\vp_2 \cdots $ of the multiparticle state $\{X\}$ constrained so that $\vec{p}_1+\vec{p}_2+\cdots = \vq$ (see fig.\ref{fig:cascade}). There is also an implicit sum over helicity states of the fermionic fields.
The coefficients $A_\Phi;A_I;A_F$ are the amplitudes of the initial, intermediate and final states respectively, $\alpha = e,\mu$ are the charged leptons (we are considering either $\pi$ or $K$ decay but $\tau$ decay can be considered along the same lines as described below), each $\alpha$ represents a different decay channel for the pseudoscalar meson $\Phi$. The processes that lead to the state (\ref{intstate})  to second order in the interaction(s) are depicted in fig.(\ref{fig:cascade}), the dots stand for higher order processes, each vertex in the diagram (\ref{fig:cascade}) corresponds to one power of the couplings in $H_I$, either at the production or decay vertices.

In what follows we distinguish the labels for the heavy sterile neutrinos as $h=1,2$, which \emph{should not} be confused with the active-like neutrinos, simply labeled as $\underline{a}$ without further   specification.

Unitary time evolution with the initial condition $A_\Phi(\vk,0)=1$ implies
\be |A_\Phi(\vk,t)|^2   + \sum_{\alpha;\vq;i=a,h}\,|A^{\alpha\, i}_{I}(\vk,\vq;t)|^2 +   \sum_{\alpha;\vq;\{X\};\{\vec{p} \}_X} |A^{\alpha\,X}_{F}(\vk,\vq,\{\vp\}_X;t)|^2 + \cdots = 1 \,. \label{unitime}\ee which has been explicitly confirmed in general in ref.\cite{cascade} and in particular for the case of single sterile neutrinos in ref.\cite{mio}.

  We introduce the following notation,
\bea && E_\Phi \equiv E_{\Phi}(k)~~;~~E^i_I \equiv E_\alpha(|\vk-\vq|)+E_{i}(q)~~;~~ i = a, h\label{energies} \\&& E^X_F \equiv E_\alpha(|\vk-\vq|)+ E^X~~;~~E^X \equiv E_{X_1}(p_1)+E_{X_2}(p_2)+\cdots \label{defenergies}\\
&& \langle \nu_{i,\vq};\,L^\alpha_{\vk-\vq} |H_I(t)| \Phi_{\vk} \rangle \equiv M^{\alpha \,i}_\mathcal{P}(\vk,\vq)\,e^{-i(E_\Phi-E^i_I)t}      \label{Mpi}\\
&& \langle L^\alpha_{\vk-\vq}\,;\, \{X\} |H_I(t)|  \nu_{h,\vq};\,L^\alpha_{\vk-\vq} \rangle  \equiv M^{h\,X}_\mathcal{D}(\vk,\vq,\vp) \,e^{-i (E^h_I-E^X_F)t}     \label{Mdj}\eea where $E_\Phi(k);E_i(q);E_\alpha(|\vk-\vq|)$ are the single particle energies for the quanta of the respective fields and $E^X$ is the energy of the   multi-particle state with the set of momenta $\{\vp\}_X$. The matrix elements $M_\mathcal{P},M_\mathcal{D}$ refer to production ($\mathcal{P}$) and decay ($\mathcal{D}$) vertices.


 For example, for the specific production vertex described by (\ref{Hintfi}) we find,
 \be M^{\alpha\,i}_\mathcal{P}(\vk,\vq;s,s') =  U_{\alpha i}\,   {F_\Phi} \, \frac{ \overline{\mathcal{U}}_{\alpha,s}(\vk-\vq)\,\gamma^\mu \,(1-\gamma^5)\, \mathcal{V}_{i,s'}(\vq)\, k_\mu }{\sqrt{32\,V\,\,E_\Phi(k)E_\alpha(|\vk-\vq|)E_i(q)}} ~~;~~i=a,h \label{prodmtx} \ee where $ \overline{\mathcal{U}}_{\alpha,s}(\vk-\vq);\mathcal{V}_{i,s'}(\vq)$ are the Dirac spinors for the charged lepton $\alpha$ and neutrino $i=a,h$, and the labels $s,s'$ refer to helicity states and will be suppressed in what follows. If neutrinos are Majorana,
  it follows that
 \be \mathcal{V}_{i,s'}(\vq) \rightarrow \mathcal{U}^{\,c}_{i,s'}(-\vq)\,. \label{majo}\ee


The counterterm in the interaction Hamiltonian $H_{ct}$ yields the matrix elements
\be \langle \nu_{h,\vq}|H_{ct}|\nu_{h',\vq}\rangle = \delta\, \mathcal{E}_{h h'} =  \delta \, \mathcal{E}^*_{h' h} \label{counter} \ee and renormalizes the masses by subtracting the hermitian parts of the self-energies as discussed in detail below. The second equality in (\ref{counter}) is a consequence of hermiticity of the interaction Hamiltonian.

To simplify notation    we  suppress  the momentum arguments of the amplitudes, energies and matrix elements, they are displayed explicitly in the expansion (\ref{intstate}) and the definitions (\ref{energies},\ref{defenergies},\ref{Mpi},\ref{Mdj}) respectively.

The time evolution of the amplitudes $A_\Phi; A^{\alpha i}_I;A^{\alpha X}_F$ is obtained from the Schroedinger equation (\ref{inttimeevol}) by projecting onto the Fock states, namely with the interaction picture state written as (\ref{ip}) it follows that
 \be \dot{A}_m(t) = -i \sum_{n}\langle m|H_I(t)|n\rangle\,A_n(t)  = -i \sum_{n} \mathcal{M}_{mn}\, e^{i(E_m-E_n)t}\,A_n(t) \,. \label{eqamps1}\ee where we have used that the  matrix elements are of the form
\be
  \langle m | H_I(t) | n \rangle = e^{i(E_m-E_n)t}\,\mathcal{M}_{mn}~~;~~ \mathcal{M}_{mn} =\langle m|H_I(0)|n\rangle \,, \label{mtxele}
\ee the relevant matrix elements are given by eqns. (\ref{Mpi},\ref{Mdj}).

\vspace{2mm}

Using eqn. (\ref{eqamps1}) we obtain the following equations
\bea  \dot{A}_{\Phi}(t) && = -i \sum_{\alpha,\vq,a} {M^{\alpha a}_\mathcal{P}}^*\,e^{i(E_\Phi-E^a_I)t}\,A^{\alpha a}_I(t) \nonumber \\&& ~~  -i \sum_{\alpha,\vq,h=1,2} {M^{\alpha\,h}_\mathcal{P}}^*\,e^{i(E_\Phi-E^h_I)t}\,A^{\alpha\,h}_I(t)
  ~;~ A_{\Phi}(0)=1  \label{dotafi1s}\\
 \dot{A}^{\alpha a}_I(t) && = -i\,e^{-i(E_\Phi-E^a_I)t}\,{M^{\alpha a}_\mathcal{P}} \, A_{\Phi}(t)~~;~~{A^{\alpha a}_I}(0)=0 ~~(\mathrm{active})\label{dotaI1a} \\
  \dot{A}^{\alpha h}_I(t) && = -i\,e^{-i(E_\Phi-E^h_I)t}\,{M^{\alpha h}_\mathcal{P}} \, A_{\Phi}(t) -i \sum_{h'=1,2} \delta \mathcal{E}_{hh'} \,e^{i(E_h-E_{h'})t}\,  {A}^{\alpha h'}_I(t) \nonumber \\ &&- i\sum_{\{X\};\{\vp\}_X} {M^{h\,X}_\mathcal{D}}^* \,e^{-i(E^X_F-E^h_I)t}\,A^{\alpha\,X}_F(t)  ~;~ {A^{\alpha\,h}_I}(0)=0 ~,~ h=1,2 ~~(\mathrm{sterile})\label{dotaI1s} \\
  \dot{A}^{\alpha X}_F (t) && = -i \sum_{h=1,2}  {M^{h\,X}_\mathcal{D}}  \,e^{i(E^X_F-E^h_I)t}\,A^{\alpha\,h}_I(t)~~;~~{A^{\alpha\,X} _F}(0)=0\,.\label{dotaF1s}\eea

  The higher order terms in the expansion of the quantum state, represented by the dots in (\ref{intstate}) lead to higher order terms in the hierarchy of equations. The label $\alpha$ in $A_I,A_F$ refer to the fact that the (charged) lepton $\alpha$ is entangled with the intermediate neutrino and final state and the kinematics of the production and decay depend on its mass.

   In ref.\cite{cascade} it is shown  that truncating the hierarchy at the order displayed above and solving the coupled set of equations provides a non-perturbative real time resummation of Dyson-type self-energy diagrams with self-energy corrections up to second order in the interactions. In appendix (\ref{sec:props}) we provide a similar analysis for the case of mixing considered here, and establish a correspondence with the self-energy treatment in refs.\cite{pilares,pila1,pila2}.

  The three terms on the right hand side in eqn. (\ref{dotaI1s}) have a clear interpretation: the first term describes the build up of the amplitude from the decay of the parent meson, the second term is the counterterm (see eqn. (\ref{counter})) and the third term describes the decay of the heavy steriles into the final states.

  The solution of the set of equations (\ref{dotafi1s}-\ref{dotaF1s})) proceeds from the bottom up. The solution of  (\ref{dotaF1s}) is
  \be  {A}^{\alpha X}_F (t)  = -i\int_0^t\, \Bigg\{{M^{1\,X}_\mathcal{D}}\,e^{i(E^X_F-E^1_I)t}\,A^{\alpha 1}_I(t')+ {M^{2\,X}_\mathcal{D}}\,e^{i(E^X_F-E^2_I)t}\,A^{\alpha 2}_I(t') \Bigg\} \,dt' \,.\label{AFXsol}\ee Introducing this solution into eqns. (\ref{dotaI1s}) we obtain
  \bea  \dot{A}^{\alpha 1}_I(t) && =  -i\,e^{-i(E_\Phi-E^1_I)t}\,{M^{\alpha 1}_\mathcal{P}} \, A_{\Phi}(t)  - i\delta\,\mathcal{E}_{11}\,A^{\alpha 1}_I(t) - i\delta\,\mathcal{E}_{12}\,e^{i(E_1-E_2)t}\,A^{\alpha 2}_I(t) \nonumber \\ &&
    -  \sum_{\{X\};\{\vp\}_X} \int^t_0\Bigg\{ |{M^{1 X}_\mathcal{D}}|^2 \,e^{-i(E^X_F-E^1_I)(t-t')}\,A^{\alpha 1}_I(t') \nonumber \\ && +   {M^{1 X}_\mathcal{D}}^*\,{M^{2 X}_\mathcal{D}}\,e^{i(E_1-E_2)t} \,e^{-i(E^X_F-E^2_I)(t-t')}\,A^{\alpha 2}_I(t')\Bigg\}\,dt' \label{eqnAI1}\eea
   \bea  \dot{A}^{\alpha 2}_I(t) && =  -i\,e^{-i(E_\Phi-E^2_I)t}\,{M^{\alpha 2}_\mathcal{P}} \, A_{\Phi}(t)  - i\delta\,\mathcal{E}_{22}\,A^{\alpha 2}_I(t) - i\delta\,\mathcal{E}_{21}\,e^{i(E_2-E_1)t}\,A^{\alpha 1}_I(t) \nonumber \\ && -    \sum_{\{X\};\{\vp\}_X} \int^t_0\Bigg\{ |{M^{2 X}_\mathcal{D}}|^2 \,e^{-i(E^X_F-E^2_I)(t-t')}\,A^{\alpha 2}_I(t')  \nonumber \\ && +   {M^{2 X}_\mathcal{D}}^*\,{M^{1 X}_\mathcal{D}}\,e^{i(E_2-E_1)t} \,e^{-i(E^X_F-E^1_I)(t-t')}\,A^{\alpha 1}_I(t')\Bigg\}\,dt' \label{eqnAI2}\,.\eea

\textbf{The Wigner-Weisskopf approximation:} In solving the hierarchy of coupled equations from the bottom up, we encounter linear integro-differential equations for the coefficients, of the general form (see (\ref{eqnAI1}, \ref{eqnAI2})).

\be \dot{A}(t)+\int^t_0 \sum_{\vp}|M |^2 e^{i(E_I-E_F )(t-t')}\,A(t') dt' = I(t) \label{wweq}\ee where $I(t)$ is an inhomogeneity. These type of equations can be solved in terms of Laplace transforms (as befits an initial value problem). In ref.\cite{cascade} it is shown that the solution of  the hierarchy of equations via Laplace transform   yields a real time non-perturbative resummation of a Dyson-type self-energy diagrams and a similar proof for the case of mixing is provided in appendix (\ref{sec:props}). An alternative but equivalent method relies on that the matrix elements $M$ are typically of $\mathcal{O}(g)$ where   $g$ refers to a generic coupling in $H_I$\cite{cascade}. Therefore in perturbation theory the amplitudes evolve \emph{slowly} in time since $\dot{A} \propto g^2 A$ suggesting an expansion in \emph{derivatives}.
 This is implemented as follows\cite{cascade,mio}, consider
\be \label{wzero}
W_0(t,t') = \sum_{\vp} |M|^2 \int^{t'}_{0} dt'' e^{-i(E_I-E_F )(t-t'')}
\ee which has the properties
\be
\frac{d}{dt'} W_0(t,t') =  \sum_{\vp} |M|^2 e^{-i(E_I-E_F )(t-t')}  \sim \mathcal{O}(g^2)~~;~~ W_0(t,0) = 0 \,. \label{propy}
\ee and is the kernel of the integral term in (\ref{wweq}).  An integration by parts in (\ref{wweq}) yields

\be
\int^t_0 dt' \frac{d}{dt'} W_0(t,t') A (t') = W_0(t,t) A (t) - \int_0^t dt' \dot{A} (t') W_0(t,t') \label{deri}
\ee   From the amplitude equations it follows that  $\dot{A} \propto g^2\,A$ and $W_0 \propto g^2$, therefore the second term on the right hand side in (\ref{deri}) is $\propto g^4$ and can be neglected to leading order $\mathcal{O}(g^2)$ which is consistent with the order at which the hierarchy is truncated. This procedure can be repeated systematically, producing higher order derivatives, which are in turn higher order in $g^2$ providing a systematic quantum field theoretical generalization of the Wigner-Weisskopf method ubiquitous in the treatment of neutral meson mixing\cite{kabir,bigi,lavoura}.

The Wigner-Weisskopf approximation is the  leading order in the coupling(s) and consists in keeping the first term in (\ref{deri}) and taking the long time limit,
 \be W_0(t,t) \rightarrow \sum_{\vp} |M|^2  \,\int^{t\rightarrow \infty}_0 e^{i(E_I-E_F +i \epsilon )(t-t'')} dt'' = i \sum_{\vp} \frac{|M|^2}{(E_I-E_F +i \epsilon )} \label{wwapx}\ee where $\epsilon \rightarrow 0^+$ is a convergence factor for the long time limit.

 A more detailed analysis of the long time limit presented in refs.\cite{boyrich,cascade} allows to extract the contribution from wave function renormalization, we will not pursue this contribution   here as it is  not directly relevant   to the time evolution and oscillations which is the focus of this study.

In ref.\cite{cascade} it is shown explicitly that this approximation is indeed equivalent to the exact solution via Laplace transform in the weak coupling and long time limit, where the Laplace transform is dominated by a narrow Breit-Wigner resonance  in the Dyson-resummed propagator. The generalization of this equivalence to the case of mixing is discussed in appendix (\ref{sec:props}).

 In the Wigner-Weisskopf  approximation   up to second order in $H_I$,  we obtain
\bea  \dot{A}^{\alpha 1}_I(t)+i  \Sigma_{11}\,A^{\alpha 1}_I(t) + i \Sigma_{12} \,A^{\alpha 2}_I(t)\,e^{i(E_1-E_2)t} & = &  -i\,e^{-i(E_\Phi-E^1_I)t}\,{M^{\alpha 1}_\mathcal{P}} \, A_{\Phi}(t) \label{dotA1ww}   \\
\dot{A}^{\alpha 2}_I(t)+i  \Sigma_{22} \,A^{\alpha 2}_I(t) + i \Sigma_{21} \,A^{\alpha 1}_I(t)\,e^{i(E_2-E_1)t} & = &  -i\,e^{-i(E_\Phi-E^2_I)t}\,{M^{\alpha 1}_\mathcal{P}} \, A_{\Phi}(t)  \label{dotA2ww}    \eea

The oscillatory factors $e^{\pm i(E_1-E_2 )t}$ in (\ref{dotA1ww},\ref{dotA2ww}) can be absorbed by defining
\be \mathcal{A}^{h}(t) \equiv e^{-iE_{h}  \, t}\,A^{\alpha h}_I(t)~~;~~h=1,2 \,  \label{defiA}\ee leading to  the following matrix equations for these amplitudes

\be \frac{d}{dt}\Big(
                       \begin{array}{c}
                         \mathcal{A}^{1}(t) \\
                         \mathcal{A}^{2}(t) \\
                       \end{array}
                    \Big) +i \mathds{H}\,\Big(
                       \begin{array}{c}
                         \mathcal{A}^{1}(t) \\
                         \mathcal{A}^{2}(t) \\
                       \end{array}
                    \Big) = -i\,e^{i(E_\alpha-E_\Phi)t}\,A_\Phi(t)\,\Big(\begin{array}{c}
                                                                                   M^{\alpha 1}_\mathcal{P} \\
                                                                                   M^{\alpha 2}_\mathcal{P}
                                                                                 \end{array}
                                     \Big) \label{mtxeqn}\ee

where the ``effective Hamiltonian''
 \be \mathds{H} \equiv \Bigg( \begin{array}{cc}
                                      H_{11}  & H_{12}   \\
                                      H_{21}  & H_{22}
                                    \end{array} \Bigg)
 = \Bigg( \begin{array}{cc}
                                      E_1+\Sigma_{11}  & \Sigma_{12}   \\
                                      \Sigma_{21}  & E_2+\Sigma_{22}
                                    \end{array} \Bigg)\,.\label{massmtx} \ee

 The right hand side of (\ref{mtxeqn}) describes the production from $\Phi-$ decay.

                                    The matrix elements are given by

\bea \Sigma_{11} & = &  \sum_{\{X\};\{\vp\}_X} \frac{|M^{1X}_\mathcal{D}|^2}{E_1-E^X+i\epsilon}+\delta\,\mathcal{E}_{11} \equiv \Delta E_{11}+\delta\,\mathcal{E}_{11} -i \frac{\Gamma_{11}}{2} \label{sig11}\\
\Sigma_{22} & = &  \sum_{\{X\};\{\vp\}_X} \frac{|M^{2X}_\mathcal{D}|^2}{E_2-E^X+i\epsilon}+\delta\,\mathcal{E}_{22} \equiv \Delta E_{22}+\delta\,\mathcal{E}_{22} -i \frac{\Gamma_{22}}{2} \label{sig22}\\
\Sigma_{12} & = &  \sum_{\{X\};\{\vp\}_X} \frac{{M^{1X}_\mathcal{D}}^* \,M^{2X}_\mathcal{D}}{E_2-E^X+i\epsilon}+\delta\,\mathcal{E}_{12} \equiv \Delta E_{12}+\delta\,\mathcal{E}_{12} -i \frac{\Gamma_{12}}{2} \label{sig12}   \\
\Sigma_{21} & = &  \sum_{\{X\};\{\vp\}_X} \frac{ M^{1X}_\mathcal{D}\,{M^{2X}_\mathcal{D}}^*}{E_1-E^X+i\epsilon}+\delta\,\mathcal{E}_{21}  \equiv \Delta E_{21}+\delta\,\mathcal{E}_{21} -i \frac{\Gamma_{21}}{2} \label{sig21} \eea

where
 \be \Delta E_{ij} = \sum_{\{X\};\{\vp\}_X} \mathcal{P} \Bigg( \frac{ {M^{iX}_\mathcal{D}}^* \,M^{jX}_\mathcal{D}}{E_j-E^X}\Bigg)   \label{deltaEij}\ee and
\be \Gamma_{ij} = 2\pi \sum_{\{X\};\{\vp\}_X}{M^{iX}_\mathcal{D}}^*\,M^{jX}_\mathcal{D}\,\delta(E_j-E^X) \,.\label{gammaij}\ee where we used $E^h_I-E^X_F= E_h -E^X$ from eqns. (\ref{energies},\ref{defenergies}). In the expressions above the sum over $\{X\}$ refers to sum over all decay channels and $\{\vec{p}\}_X$ refer to the sum over the momenta for a fixed channel.

\begin{figure}[h!]
 \begin{center}
 \includegraphics[height=3in,width=3in,keepaspectratio=true]{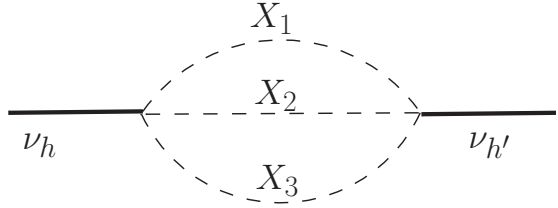}
 \caption{Self-energy that mixes $\nu_h,\nu_{h'}$ for the case of a common   three body decay channel $\nu_h;\nu_{h'} \rightarrow X_1 X_2 X_3$.   }
  \label{fig:selfie}
 \end{center}
 \end{figure}

The off-diagonal matrix elements $\Sigma_{12},\Sigma_{21}$ can be understood from the fact that the interaction Hamiltonian has non-vanishing matrix elements between the two sterile neutrinos and the \emph{same final state}. For the case of a three body common decay channel, the self energy that mixes $\nu_h, \nu_{h'}$ is depicted  in fig. (\ref{fig:selfie}), the imaginary part of this self-energy yields the widths $\Gamma_{ij}$ in eqn. (\ref{gammaij}).

As discussed in   in refs.\cite{cascade,mio}, the quantum field theoretical Wigner- Weisskopf approximation is equivalent to a Dyson resummation of Feynman diagrams and a Breit-Wigner approximation (complex pole) of the Dyson resummed propagator. This equivalence is discussed in appendix (\ref{sec:props}) and is  confirmed by the results of ref.\cite{pilares} where mixing has been studied in terms of self-energy corrections obtained from Feynman diagrams and compared to the ``effective Hamiltonian'' description within a different context. In particular, the renormalization corrections and decay widths are exactly those obtained from a Breit-Wigner approximation to the full propagator with self-energy corrections obtained from Feynman diagrams\cite{pilares}.


We emphasize that if the heavy sterile neutrinos are not exactly degenerate, namely if $E_1 \neq E_2$ then  $\Delta E_{ij}\neq (\Delta E_{ji})^*$. As a consequence of the hermiticity of the counterterm Hamiltonian it follows that $\delta\,\mathcal{E}_{ij}= (\delta\,\mathcal{E}_{ji})^*$, therefore the counterterms \emph{cannot} completely cancel the  real part of the self energy corrections  $\Delta E_{ij}$.

It is convenient to introduce the following quantities:
\bea \overline{E} & = & \frac{1}{2}\big(E_1+E_2) ~~;~~ \Delta = \frac{1}{2}\big(E_1-E_2) \label{sumdifE}\\
\overline{\Sigma} & = & \frac{1}{2}\big(\Sigma_{11}+\Sigma_{22}\big)~~;~~\sigma = \frac{1}{2}\big(\Sigma_{11}-\Sigma_{22}\big)\,, \label{sumdifvarepsi}\eea in terms of which the complex eigenvalues of $\mathds{H}$ are
\be \lambda^\pm = (\overline{E}+\overline{\Sigma})\pm \Big[(\Delta+\sigma)^2+\Sigma_{12}\,\Sigma_{21} \Big]^\frac{1}{2} \equiv E^{\pm} -i\,\frac{\Gamma^\pm}{2}  \,,\label{lambdascomp}\ee where $E^{\pm}$ and $\Gamma^{\pm}$ are real corresponding to the energy and decay width of the propagating modes.

Consider now the eigenvalue problem
\be \mathds{H}\,\Big(
                       \begin{array}{c}
                         \alpha^{\pm}_1 \\
                         \alpha^{\pm}_2 \\
                       \end{array}
                    \Big) = \lambda^{\pm} \Big(
                       \begin{array}{c}
                         \alpha^{\pm}_1 \\
                         \alpha^{\pm}_2 \\
                       \end{array}
                    \Big)   \label{eigen}\ee and  the matrices
 \be \mathcal{U}^{-1} = \Bigg(\begin{array}{cc}
                      \alpha^+_1 & \alpha^-_1 \\
                      \alpha^+_2 & \alpha^-_2
                    \end{array}
  \Bigg) ~~;~~ \mathcal{U} = \frac{1}{\big(\alpha^+_1\alpha^-_2-\alpha^+_2\alpha^-_1\big)} ~~ \Bigg(\begin{array}{cc}
                      \alpha^-_2 & -\alpha^-_1 \\
                      -\alpha^+_2 & \alpha^+_1
                    \end{array}
  \Bigg) \label{Umtx}\ee from which it follows that
  \be \mathcal{U} ~\mathds{H} ~ \mathcal{U}^{-1} = \Bigg( \begin{array}{cc}
                                     \lambda^+ & 0 \\
                                     0 & \lambda^-
                                   \end{array}
  \Bigg) \,.\label{Mdiag} \ee

  Therefore, defining
  \be   \Big(
                       \begin{array}{c}
                         \mathcal{A}^{1}(t) \\
                         \mathcal{A}^{2}(t) \\
                       \end{array}
                    \Big) = \mathcal{U}^{-1} ~\Big(
                       \begin{array}{c}
                          {V}^{+}(t) \\
                          {V}^{-}(t) \\
                       \end{array}
                    \Big)\, \label{Vdef} \ee and right-multiplying (\ref{mtxeqn}) by $\mathcal{U}$ and using (\ref{Mdiag}) we find

\be \frac{d}{dt}~ \Big(
                       \begin{array}{c}
                         {V}^{+}(t) \\
                          {V}^{-}(t) \\
                       \end{array}
                    \Big) +i  \Bigg( \begin{array}{cc}
                                     \lambda^+ & 0 \\
                                     0 & \lambda^-
                                   \end{array}
  \Bigg)~     \Big(
                       \begin{array}{c}
                          {V}^{+}(t) \\
                           {V}^{-}(t) \\
                       \end{array}
                    \Big)  = -i\,e^{i(E_\alpha-E_\Phi)t}\,A_\Phi(t)\,\Big(\begin{array}{c}
                                                                                  \widetilde{ M}^{\alpha+}_\mathcal{P} \\
                                                                                   \widetilde{M}^{\alpha -}_\mathcal{P}
                                                                                 \end{array}
                                     \Big) ~~;~~                                                  {V}^{\pm}(0) =0
                          \label{Veqn} \ee

with
\be \Big(\begin{array}{c}
                                                                                   \widetilde{M}^{\alpha +}_\mathcal{P} \\
                                                                                   \widetilde{M}^{\alpha -}_\mathcal{P}
                                                                                 \end{array}
                                     \Big) = \mathcal{U} ~ \Big(\begin{array}{c}
                                                                                   {M^{\alpha 1}_\mathcal{P} }  \\
                                                                                   {M^{\alpha 2}_\mathcal{P}}
                                                                                 \end{array}
                                     \Big) \,.  \label{inM}\ee

The solutions are
\be   V^{\pm}(t) = -i\,\widetilde{ M}^{\alpha \pm}_\mathcal{P}\, ~e^{-i\lambda^{\pm}t}~\int^t_0 e^{ i(\lambda^{\pm}+E_\alpha-E_\Phi)t'} ~A_\Phi(t')\,dt'\,, \label{solV}\ee
and from the relation (\ref{Vdef}) we obtain
\be \mathcal{A}^{1}(t) = \alpha^+_1\,  V^+(t)+\alpha^-_1 \,  V^-(t)~~;~~ \mathcal{A}^{2}(t) = \alpha^+_2  V^+(t)+\alpha^-_2\,  V^-(t)  \,. \label{finA} \ee Full expressions for the products $ \alpha^{\pm}_{j} \widetilde{ M}^{\alpha \pm}_\mathcal{P}$ are given in appendix (\ref{sec:identities}) where it is recognized that these products are independent of the normalization of the eigenvectors of $\mathds{H}$.

 The solution of (\ref{dotaI1a}) is
\be  {A^{\alpha a}_I}(t) = -i\,{M^{\alpha a}_\mathcal{P}}^*\, \int^t_0e^{-i(E_\Phi-E^a_I)t'}\,  A_{\Phi}(t') \,dt' \,, \label{dotaI1asolu} \ee
 we now insert the  solutions (\ref{finA},\ref{dotaI1asolu}) into the evolution equation for $A_\Phi(t)$ (\ref{dotafi1s}), using the definitions (\ref{defenergies},\ref{defiA}) we find
\bea \dot{A}_\Phi(t) & = &   -     \sum_{\alpha;\vq} \int^t_0 \Bigg\{\overline{M}^{\alpha +}_\mathcal{P}\,\widetilde{M}^{\alpha +}_\mathcal{P} ~e^{ -i(\lambda^{+}+E_\alpha-E_\Phi)(t-t')} + \overline{M}^{\alpha -}_\mathcal{P}\,\widetilde{M}^{\alpha -}_\mathcal{P} ~e^{ -i(\lambda^{-}+E_\alpha-E_\Phi)(t-t')}   \Bigg\} A_\Phi(t')~dt' \nonumber \\
& - &  \sum_{\alpha;\vq;a} \int^t_0 \Bigg\{|M^{\alpha a}_\mathcal{P}|^2 e^{-i(E_a+E_\alpha-E_\Phi)(t-t')}\,A_\Phi(t')\Bigg\}~dt'
  \,, \label{Phieq2}\eea where
\be \overline{{M}}^{\alpha \pm}_\mathcal{P} = \alpha^{\pm}_1 {M^{\alpha 1}_\mathcal{P}}^*  + \alpha^{\pm}_2 {M^{\alpha 2}_\mathcal{P}}^* \,. \label{tilMp}\ee

The first line in eqn. (\ref{Phieq2}) is the contribution from the  intermediate heavy sterile states, and the second line is the contribution from the active neutrinos.

Implementing the Wigner-Weisskopf approximation and taking the long time limit (no convergence factor is needed  in the first sum because $\lambda^{\pm}$ feature a negative imaginary part arising from the decay of the intermediate state) this evolution equation simplifies to
\be \dot{A}_\Phi(t) + i~\mathcal{E}_\Phi\,A_\Phi(t) = 0 ~~;~~ A_\Phi(0) =1 \label{wwAfi}\ee where
\bea   \mathcal{E}_\Phi    & \equiv &    \Delta E_{\Phi} - i~\frac{\Gamma_\Phi}{2} = \sum_{\alpha;\vq;a} \frac{|M^{\alpha a}_\mathcal{P}|^2}{E_\Phi-E_a-E_\alpha +i \epsilon } \nonumber \\
 & +&
\sum_{\alpha;\vq} \Bigg\{ \frac{\overline{M}^{\alpha +}_\mathcal{P}\,\widetilde{M}^{\alpha +}_\mathcal{P} }{\big(E_\Phi-E_\alpha -E^+ +\frac{i}{2}\Gamma^+\big)} +  \frac{\overline{M}^{\alpha -}_\mathcal{P}\,\widetilde{M}^{\alpha -}_\mathcal{P} }{\big(E_\Phi-E_\alpha -E^- +\frac{i}{2}\Gamma^-\big)}  \Bigg\}  \label{defEfi}\eea
 with $\Delta E_\Phi~;~\Gamma_\Phi$ real, leading to
\be A_\Phi(t) = e^{-i~\Delta E_\Phi\,t} ~~ e^{-\frac{\Gamma_\Phi}{2}\,t} \,.\label{finAfi}\ee $\Delta E_\Phi$ will be absorbed into a  renormalization of the single meson energy, namely $E_\Phi+\Delta E_\Phi \rightarrow E_\Phi$ (from now on $E_\Phi$ denotes the renormalized single particle energy) and $\Gamma_\Phi$ is the \emph{total decay width} of the parent meson.

It only remains to introduce the result (\ref{finAfi}) into (\ref{solV})   to obtain the time evolution of all the amplitudes. We find
\be V^{\pm}(t) = \widetilde{ M}^{\alpha \pm}_\mathcal{P}\, \frac{\Big[e^{-i(E_\Phi-E_\alpha-i\frac{\Gamma_\Phi}{2})t}-e^{-i(E^\pm - i\,\frac{\Gamma^\pm}{2})\,t} \Big]}{\Big[E_\Phi-E_\alpha-E^\pm-\frac{i}{2}\,(\Gamma_\Phi-\Gamma^\pm)\Big]}\,.\label{Vpmsolu}\ee
Using the definition (\ref{defiA}) and inserting the results (\ref{finA},\ref{Vpmsolu}) into (\ref{dotaF1s}) we find for the final state amplitude
\bea A^X_F(t) & = & \frac{ \Big(\alpha^+_1 \widetilde{M}^{\alpha +}_\mathcal{P}\, M^{1X}_\mathcal{D}+ \alpha^+_2 \widetilde{M}^{\alpha +}_\mathcal{P}\, M^{2X}_\mathcal{D}\Big)}{\Big[E_\Phi-E_\alpha-E^+ -\frac{i}{2}\,(\Gamma_\Phi-\Gamma^+) \Big]}\,\Bigg\{\frac{\Big[e^{-i(E_\Phi-E^X_F-i\frac{\Gamma_\Phi}{2})t}-1 \Big]}{\Big[ E_\Phi-E^X_F-i\frac{\Gamma_\Phi}{2}\Big]}-
\frac{\Big[e^{-i(E^+ -E^X-i\frac{\Gamma^+}{2})t}-1 \Big]}{\Big[ E^+  -E^X-i\frac{\Gamma^+}{2}\Big]} \Bigg\} \nonumber \\ & + &
 \frac{ \Big(\alpha^-_1 \widetilde{M}^{\alpha -}_\mathcal{P}\, M^{1X}_\mathcal{D}+ \alpha^-_2 \widetilde{M}^{\alpha -}_\mathcal{P}\, M^{2X}_\mathcal{D}\Big)}{\Big[E_\Phi-E_\alpha-E^- -\frac{i}{2}\,(\Gamma_\Phi-\Gamma^-) \Big]}\,\Bigg\{\frac{\Big[e^{-i(E_\Phi-E^X_F-i\frac{\Gamma_\Phi}{2})t}-1 \Big]}{\Big[ E_\Phi-E^X_F-i\frac{\Gamma_\Phi}{2}\Big]}-
\frac{\Big[e^{-i(E^-  -E^X-i\frac{\Gamma^-}{2})t}-1 \Big]}{\Big[ E^- -E^X-i\frac{\Gamma^-}{2}\Big]} \Bigg\}\,.  \nonumber \\ \label{Afinat}   \eea

In the probability of detecting the final state $| A^X_F(t) |^2$ the interference between the terms with $e^{-i(E^\pm -i \frac{\Gamma^\pm}{2})t}$ leads to oscillations. These will be studied in section (\ref{sec:oscillations}) below.

Going back to the Schroedinger picture with $|\Psi(t)\rangle_S = e^{-iH_0t}\,|\Psi(t)\rangle_I $ we obtain
\bea
&& |\Psi(\vk,t)\rangle_S   =   e^{-iE_\Phi t}\,e^{-\frac{\Gamma_\Phi}{2}t}\,\big|\Phi_{\vec{k}}\rangle +\sum_{\alpha;\vq; a}\,e^{-iE^\alpha_I t}\,  {A}^{\alpha\, a}_{I}(\vk,\vq;t)\,\big|\nu_{a,\vq};\,L^\alpha_{\vk-\vq} \rangle \nonumber \\
&+&   \sum_{\alpha;\vq;h}\,e^{-iE_\alpha t}\, \mathcal{A}^{ h} (\vk,\vq;t)\,\big|\nu_{h,\vq};\,L^\alpha_{\vk-\vq} \rangle +  \sum_{\alpha;\vq;\{X\};\{\vec{p} \}_X} e^{-iE^X_F t} A^{\alpha\,X}_{F}(\vk,\vq,\{\vp\}_X;t)\,\big|L^\alpha_{\vk-\vq}\,; \{X\} \rangle  +\cdots \nonumber \\
  \label{Psischo}
\eea

Using the result (\ref{finA}) and the definition (\ref{defiA}) we note that we can write  in the second term in (\ref{Psischo})
\be \sum_{h=1,2} \mathcal{A}^h (t)\,\big|\nu_h\rangle = V^+(t)|\nu^+\rangle + V^-(t)|\nu^-\rangle ~~;~~ |\nu^{\pm}\rangle = \alpha^\pm_1  |\nu_1\rangle+\alpha^\pm_2  |\nu_2\rangle \label{steripm}\ee namely the states $|\nu^\pm \rangle$ are coherent superpositions of the mass eigenstates of the unperturbed Hamiltonian, in particular under the assumption that $\Gamma_\Phi \gg \Gamma^\pm$   it follows that for time scales $1/\Gamma_\Phi \ll t \lesssim 1/\Gamma_{\pm}$
\be V^\pm(t) = C_{\pm}~ e^{-iE^\pm t}\,e^{-\Gamma^\pm t} \label{expospm}\ee where the normalization constants
  \be C_\pm =  -  \frac{\widetilde{ M}^{\alpha \pm}_\mathcal{P}}{\Big[E_\Phi-E_\alpha-E^\pm-\frac{i}{2}\,(\Gamma_\Phi-\Gamma^\pm)\Big]}\label{Cis} \ee
       reflect the Lorentzian distribution from the decay of the parent meson. However, because $\mathds{H}$ is non-hermitian the  states $|\nu^\pm\rangle$ are not orthogonal, namely $\langle \nu^+|\nu^-\rangle = (\alpha^{+}_1  )^{*}\, (\alpha^-_1  )+ (\alpha^{+}_2 )^{*}(\alpha^-_2  ) \neq 0$.

\vspace{2mm}

\textbf{Comparison to neutral meson mixing:}

\vspace{2mm}

The evolution equations for the amplitudes of the intermediate state (\ref{mtxeqn}) in terms of an ``effective Hamiltonian'' (\ref{massmtx})   are similar to the case of neutral meson mixing but with noteworthy differences:

\begin{itemize}
\item   The inhomogeneity on the right hand side of (\ref{mtxeqn}) describes the \emph{production} of the intermediate state from the decay of the initial state. In the description of neutral meson mixing, the production stage is not included but the initial state is assumed to be a linear superposition of the unperturbed neutral mesons ($K^0,\overline{K^0}, B^0, \overline{B^0}$ etc.), and the equivalent of eqn. (\ref{mtxeqn}) is homogeneous. Since the amplitude $A_\Phi(t) \rightarrow 0$ for $t\gg 1/\Gamma_\Phi$ the production contribution vanishes   and eqn. (\ref{mtxeqn}) becomes homogeneous describing an initial value problem for the amplitudes for time scales $t\gg 1/\Gamma_\Phi$, therefore one would conclude that for $t\gg 1/\Gamma_{\Phi}$ the two cases are similar. However, it is clear from the expressions (\ref{Vpmsolu}) that in this limit, the amplitudes for the heavy sterile neutrinos \emph{are not determined from  arbitrary initial conditions}, but are determined by the Lorentzian distribution function that results from the decay of the parent particle. This is manifest in the prefactors $C_{\pm}$ in (\ref{expospm})  which are given by (\ref{Cis}) as a direct consequence of production of sterile neutrinos from the decay process, in other words, these coefficients are a manifestation of the ``memory'' of the initial state and  of the decay dynamics of the parent meson.

    The probability for finding a particular mode $\pm$ after the decay of the parent meson for $t\gg 1/\Gamma_\Phi$ is
      \be  \frac{|\widetilde{ M}^{\alpha \pm}_\mathcal{P}|^2\,e^{-\Gamma^\pm\,t}}{\Big[E_\Phi-E_\alpha-E^\pm\Big]^2+ \Big[\big(\Gamma_\Phi-\Gamma^\pm\big)/2\Big]^2 }\,, \label{proba} \ee namely the exponential decay factor multiplies a Lorentzian probability distribution  of decay products. The difference in the decay widths in the denominator has a simple interpretation: $\Gamma_\Phi$ describes the rate at which the sterile neutrinos are \emph{produced}, whereas $\Gamma^\pm$ are the rate at which they \emph{decay} into the final states so that the effective production rate is $\Gamma_\Phi - \Gamma^\pm$.

   \item Unlike   the neutral meson case under the assumption of CPT symmetry, the diagonal entries in the matrices (\ref{deltaEij},\ref{gammaij}) are \emph{not} the same. This is because the sterile neutrinos in the intermediate state are \emph{not exactly degenerate}   As a consequence of  this non-degeneracy, it also follows that $\Delta E_{ij} \neq \Delta E^*_{ji}~;~\Gamma_{ij} \neq \Gamma^*_{ji} $ unlike the case of neutral meson mixing. Therefore, as mentioned above, the counterterms $\delta\mathcal{E}_{ij}$ obeying the hermiticity condition  cannot completely cancel the self energy corrections $\Delta E_{ij}$.
        In the case of neutral meson mixing, the unperturbed (bare) masses of the meson and antimeson are the same, hence the denominators in $\Delta E_{12}, \Delta E_{21}$ are the same and $\Delta E_{ij}$ is hermitian\cite{kabir}. Indeed   the original derivation in\cite{kabir}   manifestly  uses that the meson and antimeson have the same (unperturbed) energy (mass). Allowing for different energies and  following the derivation in\cite{kabir}  the results for $\Delta E_{ij}$ obtained above follow directly.

\item The time dependent prefactors $V^\pm(t)$ are given by (\ref{Vpmsolu}), the first term $\propto e^{-i(E_\Phi-E_\alpha-i\frac{\Gamma_\Phi}{2})t}$ is a direct consequence of the \emph{production} of sterile neutrinos via the decay of the pseudoscalar meson and can be traced to the right hand side of eqn. (\ref{mtxeqn}).  \emph{If} $\Gamma_\Phi \gg \Gamma^\pm$ and for $ t \gg 1/\Gamma_\Phi$ it follows that $V^\pm(t) \propto e^{-i(E^\pm - i\,\frac{\Gamma^\pm}{2})\,t}$ which is the usual time evolution obtained from the ``effective Hamiltonian'' in   the Wigner-Weisskopf approximation for neutral meson mixing\cite{kabir,bigi,lavoura}. This is in agreement with the results of ref.\cite{cascade} wherein it was observed that if the decay rate of the parent particle is much larger than that of the intermediate resonant state, the time evolution proceeds sequentially:  the decay of the parent particle   leads to the formation of the intermediate state on a time scale much shorter than the lifetime of the intermediate resonant state, its amplitude grows initially from the production dynamics and  decays on a longer time scale.

\end{itemize}

\vspace{2mm}

\section{Oscillations in the detection of decay products.}\label{sec:oscillations}

\vspace{2mm}

Oscillations in the decay products are observationally relevant on macroscopic scales when the heavy sterile neutrinos are nearly degenerate, namely when $E_1+E_2 \gg |E_1-E_2|$. In this limit there are two important cases to consider:

\vspace{2mm}

\textbf{I:} $\mathbf{|E_1-E_2| \gg \Sigma_{ij}}$. Since $\Sigma_{ij}\propto g^2 $ where   $g$ is a typical weak coupling in the interaction Hamiltonian, we find up to second order in couplings
\be  \lambda^+ = E_1+ \Sigma_{11} + \mathcal{O}(g^4) ~~;~~ \lambda^- = E_2+\Sigma_{22} + \mathcal{O}(g^4) \label{Deltalarge}\ee the counterterms can be chosen to cancel the real parts of the self-energy so that $E_{1,2}$ are the fully renormalized (real) energies and to leading order in the couplings for this case we find
\be \lambda^+ = E_1 -\frac{i}{2}\Gamma_{11} ~~;~~\lambda^- = E_2 -\frac{i}{2}\Gamma_{22} \,. \label{dellar}\ee

\vspace{2mm}

\textbf{II:} $\mathbf{|E_1-E_2| \lesssim  \Sigma_{ij}}$. In this case the full expression for   $\lambda^\pm$ are given by (\ref{lambdascomp}) and   we can set $E_{1,2} \rightarrow \overline{E}=(E_1+E_2)/2$ to leading order in the self-energies (\ref{sig11}-\ref{sig21}). Neglecting terms of $\mathcal{O}(g^2 |\Delta|/\overline{E} \lesssim g^4)$, we find
\be \lambda^+ - \lambda^- = 2 \, \Big[(\Delta+\sigma)^2+\Sigma_{12}\,\Sigma_{21} \Big]^\frac{1}{2} \propto \mathcal{O}(g^2)\,. \label{delles} \ee where
\be \Delta E_{ij} = \sum_{\{X\};\{\vp\}_X} \mathcal{P} \Bigg( \frac{ {M^{iX}_\mathcal{D}}^* \,M^{jX}_\mathcal{D}}{\overline{E}-E^X}\Bigg)   \label{deltaEijav}\ee and
\be \Gamma_{ij} = 2\pi \sum_{\{X\};\{\vp\}_X}{M^{iX}_\mathcal{D}}^*\,M^{jX}_\mathcal{D}\,\delta(\overline{E}-E^X) \,.\label{gammaijdelless}\ee with the corollary that $(\Delta E_{ij})^*= \Delta E_{ji}~;~(\Gamma_{ij})^*=\Gamma_{ji}$. Because the counterterms obey the hermiticity conditions (see (\ref{counter})) in this case    we implement the ``on-shell'' renormalization scheme following \cite{pilares} and request that
 \be  \Delta E_{ij}+\delta\mathcal{E}_{ij}=0\,, \label{OSren} \ee where $\Delta E_{ij}$ are given by (\ref{deltaEijav}).

The probability of finding a particular final state $X$ at time $t$ is given by $|A^{\alpha X}_F(t)|^2$. Consider the ``visible'' decay of the heavy sterile neutrinos  to the \emph{common decay channel} $\{X\} = e^+ e^- \nu_a$ namely  $\nu_{h1,h2} \rightarrow e^+ e^- \nu_a$  where $\nu_a$ is an active neutrino. The number of $e^+e^-$ \emph{pairs} in this state is given by (suppressing the appropriate quantum numbers)
\be \langle \Psi(t)|b^\dagger_e b_e d^\dagger_e d_e|\Psi(t) \rangle = |A^{\alpha e^+ e^- \nu_a}_F(t)|^2 \,, \label{dilenum}\ee and the total number of $e^+ e^-$ pairs in this particular decay channel is
\be N_{e^+ e^-}(t) = \sum_{\{\vp\}_{X}}|A^{\alpha e^+ e^- \nu_a}_F(t)|^2\,. \label{totepem}\ee
The amplitude $ A^{\alpha X}_F(t) $ (\ref{Afinat}) clearly indicates that $|A^{\alpha X}_F(t)|^2$ features oscillatory  contributions  from the interference between the terms with $e^{\pm i E^\pm t}$. These interference terms will be manifest over macroscopic distances if the real part of the eigenvalues $E^\pm$ are nearly degenerate. Since the self-energies are perturbative, from the expressions (\ref{lambdascomp}) it is clear that near degeneracy of $E^{1,2}$   implies near degeneracy of $E^\pm$. It is convenient to define
\be \overline{\mathcal{E}} = \frac{1}{2}(E^+ + E^-)~~;~~ \delta = \frac{1}{2}(E^+ - E^-)~~;~~\overline{\Gamma} = \frac{1}{2}(\Gamma^+ + \Gamma^-) = \frac{1}{2}(\Gamma_{11} + \Gamma_{22}) \,, \label{defini2}\ee with $\delta \ll \overline{\mathcal{E}}$. Writing $A^{\alpha e^+ e^- \nu_a}_F(t)\equiv A^+(t)+A^-(t)$  and assuming that the matrix elements are smooth functions of the energy so that to leading order we can evaluate them at the average energy $\overline{\mathcal{E}}$ thereby neglecting terms of $\mathcal{O}(\delta/\overline{\mathcal{E}})$ we find
\be  (A^+(t))^* A^-(t) =  {\tau^+ \tau^-} \frac{4\pi^2 \, \delta(\overline{\mathcal{E}}+E_\alpha-E_\Phi)\,{\delta(\overline{\mathcal{E}}-E^X)} }{\Big[\Gamma_\Phi+\overline{\Gamma}+2i\delta\Big]~\Big[\overline{\Gamma} -i  {2\delta} \Big]} \,\, {\Big[1-e^{2i\delta t}\,e^{-\overline{\Gamma}t} \Big]} \,, \label{intfinifor}\ee where
\be \tau^\pm = \Big(\alpha^\pm_1 \widetilde{M}^{\alpha \pm}_\mathcal{P}\, M^{1X}_\mathcal{D}+ \alpha^\pm_2 \widetilde{M}^{\alpha \pm}_\mathcal{P}\, M^{2X}_\mathcal{D}\Big)\,.\label{tausdef}\ee

The details of the calculation   are given in appendix (\ref{app:deriv}).

Integration over the final state phase space ${p}_X$ and over $\overline{\mathcal{E} }$ yield the overall energy momentum conservation and fixes the average $\overline{\mathcal{E} } = E_\Phi-E_\alpha$.

These oscillations in the probability of decay products are akin to   ``quantum beats'' in the photodetection probability of radiative decays in  multilevel atomic systems\cite{meystre} and a similar phenomenon has been discussed in ref.\cite{merle} within a different context.

In the nearly degenerate case, after imposing the ``on-shell'' renormalization condition\cite{pilares} (\ref{OSren}) we find
\be \Sigma_{ij} = -i \frac{\Gamma_{ij}}{2}  \,,  \label{sigsdege}\ee and
\be \lambda^{\pm} \equiv E^{\pm}-\frac{i}{2}\Gamma^{\pm}=  \overline{E}-\frac{i}{2}(\Gamma_{11}+\Gamma_{22}) \pm \Bigg[\Big(\Delta-\frac{i}{2}(\Gamma_{11}-\Gamma_{22}) \Big)^2 -\frac{1}{4}|\Gamma_{12}|^2\Bigg]^{1/2}\,.\label{lampm}\ee

These are general results in the nearly degenerate case.

\section{An example: the ``visible'' decay channel $\nu_h \rightarrow e^+ e^- \nu_a$}\label{sec:visible}
We now study the specific example of two nearly degenerate heavy sterile neutrino with a common purely leptonic ``visible'' decay channel: $\nu_h,\nu_{h'} \rightarrow e^+ e^- \nu_a$ via a charged and or neutral current vertex, with $a$ an active-like neutrino\cite{shrock}.

\begin{figure}[h!]
 \begin{center}
 \includegraphics[height=3in,width=3in,keepaspectratio=true]{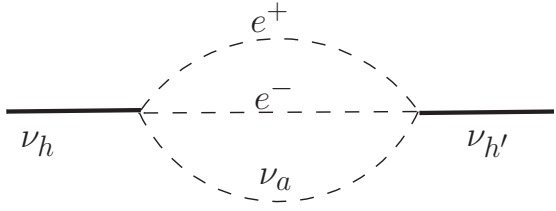}
 \caption{Self-energy with $e^+ e^- \nu_a$ in the intermediate state mixing $\nu_h,\nu_{h'}$  in the Fermi limit.   }
  \label{fig:selfvisible}
 \end{center}
 \end{figure}

In the Fermi limit the self-energy diagram that describes this common decay channel is shown in fig. (\ref{fig:selfvisible}).

   Adapting the results from ref.\cite{shrock} and neglecting corrections of order $|E^1-E^2|/(E^1+E^2)\ll 1$ we find
 \be \Gamma_{ij}(\nu_h \rightarrow e^+ e^- \nu_a) = \frac{G^2_F\, \overline{M_{h}}^{\,6}}{192 \pi^3\,\overline{E}} ~ \mathcal{H}\Big(\frac{m^2_e}{ \overline{M^2_{h}}}\Big) \,  U_{eh_i}U^*_{eh_j} ~~;~~ \overline{M_h} = \frac{1}{2}(m_{h_1}+m_{h_2})
   \label{gamahee}\ee
where\cite{shrock}
\be \mathcal{H}(x) = (1-4x^2)^\frac{1}{2}\,(1-14x-2x^2-12x^3)+24x^2(1-x^2)\ln\frac{1+(1-4x^2)^\frac{1}{2}}{1-(1-4x^2)^\frac{1}{2}} \,,\ee and to leading order we have replaced $\overline{\mathcal{E}} \rightarrow \overline{E}$. Neglecting $m_e$ it follows that\footnote{The factor $\overline{E}/\overline{M_{h}}$ is the average Lorentz factor.}
\be \Gamma_{ij}(\nu_h \rightarrow e^+ e^- \nu_a) \simeq 3.5\times  \, \Bigg[\frac{\overline{M_h}}{100\,\mathrm{MeV}}\Bigg]^5\,  \Bigg[ \frac{U_{eh_i}U^*_{eh_j} }{ 10^{-5}}\Bigg] \Bigg( \frac{\overline{M_{h}}}{\overline{E}}\Bigg) \,\mathrm{s}^{-1}\,, \label{Ghee}\ee with an extra factor of $2$ if $\nu_h$ is a Majorana neutrino.

Imposing the ``on-shell'' renormalization condition (\ref{OSren}) we find the ``effective Hamiltonian''

\be \mathds{H} =  \left(
                         \begin{array}{cc}
                           E_1-\frac{i}{2}\Gamma_{11}(\nu_h \rightarrow e^+ e^- \nu_a) & -\frac{i}{2}\Gamma_{12}(\nu_h \rightarrow e^+ e^- \nu_a) \\
                           -\frac{i}{2}\Gamma^*_{12}(\nu_h \rightarrow e^+ e^- \nu_a) &  E_2-\frac{i}{2}\Gamma_{22}(\nu_h \rightarrow e^+ e^- \nu_a) \\
                         \end{array}
                       \right)\,.
  \label{Hfina}\ee

Because the decay width is suppressed by the small neutrino mixing matrix elements, the decay vertices are expected to be displaced far from the production vertices and the space-time evolution of the sterile neutrinos becomes important. In ref.\cite{mio} these aspects were studied within the context of a single sterile neutrino  but the results are straightforwardly adapted to the present study. To address the space-time evolution a wave packet description is necessary and it is discussed in detail in ref.\cite{mio} for the case of a single sterile neutrino in a cascade decay. Consider now the nearly degenerate sterile neutrinos propagating as wave packets with nearly equal group velocities
\be v_g = \overline{p^*}/\overline{E} \label{vgroup}\ee where we have approximated $\overline{\mathcal{E}} \simeq \overline{E}$ to leading order in weak coupling and $\overline{p^*}$ is the value of the momentum determined by energy momentum conservation at the production vertex for a sterile neutrino of average energy $\overline{E}=E_\Phi-E_\alpha$. For pseudoscalar meson decaying at rest
\be v_g = \frac{ \Big[\lambda(1,\delta_{\alpha},\overline{\delta_{h}}) \Big]^\frac{1}{2}}{(1+\overline{\delta_h}-\delta_\alpha)}\,, \label{gruvel} \ee where
\be \lambda(x,y,z) = x^2+y^2+z^2-2xy-2xz-2yz~~;~~ \delta_\alpha  = \frac{m^2_{L^\alpha}}{M^2_\Phi} ~;~\overline{\delta_{h}} = \frac{\overline{M_{h}}^{\,2}}{M^2_\Phi}\,. \label{dam}\ee

Consider a detector a distance $L_d$ from the production vertex and fiducial length $\Delta L_d$, so that $|E^+-E^-|~\Delta L_d/v_g \ll 1, \overline{\Gamma}~\Delta L_d/v_g \ll 1$, then the oscillatory contribution to  the number of events   detected within the fiducial length simplifies, namely the last term in (\ref{intfinifor}) becomes (see ref.\cite{mio} for details)
\be \frac{\Big[1-e^{2i\delta t}\,e^{-\overline{\Gamma}t} \Big]}{\Big[\overline{\Gamma} -i  {2\delta} \Big]} \rightarrow e^{2i\delta L_d/v_g}\,e^{-\overline{\Gamma}L_d/v_g}\,\Delta L_d \,, \label{oscifini}\ee namely after the phase space integrations, the number of $e^+ e^-$ pairs detected within the distance $\Delta L_d$ a distance $L_d$ away from the production region is
\be N_{e^+e^-}(t)\big|_{osc} = \mathcal{N} e^{2i\delta L_d/v_g}\,e^{-\overline{\Gamma}L_d/v_g}\,\Delta L_d ~~;~~ \delta = \frac{1}{2}(E^+ - E^-)~~;~~\overline{\Gamma} = \frac{1}{2}(\Gamma^+ + \Gamma^-)  \label{oscinumee}\ee where $\mathcal{N}$ is the normalization factor arising from the phase space integrations and
\be 2\delta =  \mathrm{Re} \,\Bigg[\Big(E_1-E_2- {i}(\Gamma_{11}-\Gamma_{22}) \Big)^2 - |\Gamma_{12}|^2\Bigg]^{1/2}~~;~~ \overline{\Gamma}= \frac{1}{2}\,(\Gamma_{11}+\Gamma_{22})\,,\label{delgamvis}\ee with $\Gamma_{ij}$ given by eqn. (\ref{gamahee}).
This result for $\overline{\Gamma}$ follows from that in (\ref{oscinumee}) and (\ref{lampm}).


\subsection{Coherence aspects:}
The oscillatory behavior arising from the interference terms between the two nearly degenerate eigenstates bears many similarities with the case of oscillation and mixing of active neutrinos, but with noteworthy differences.

Oscillations in the decay products will be observed provided that $|E^+ - E^-| \gtrsim \overline{\Gamma}$ otherwise the   interference term damps out before any oscillation can occur. Furthermore, the result for the interference term (\ref{intfinifor}) has been obtained under the assumption that the difference in energies of correct eigenstates $|E^+ - E^-|$ \emph{cannot} be discriminated by the measurement. This is manifest in the derivation of eqns. (\ref{dist}) and (\ref{2fini})  in the appendix leading to the result (\ref{intfinifor}) which is obtained as a distribution integrated over a density of states (detector) that is insensitive to the energy difference. If the detector (final density of states) \emph{can} discriminate between the energy eigenstates with a resolution smaller than the widths, the narrow width approximation to each Lorentzian yields a product   $\propto \delta(\mathcal{E}-\delta)\,\delta(\mathcal{E}+\delta)$ which vanishes and the interference and quantum beats will be suppressed as the measurement is effectively projecting on a particular energy eigenstate. This is similar to the case of active   neutrino oscillations when the neutrino mass eigenstates are produced in the decay of a parent meson whose decay width determines the energy resolution as analyzed in refs.\cite{lou,boycohe} and discussed further in ref.\cite{kopp}.

The analysis leading to the result (\ref{oscinumee}) made use of a wave packet description of the space time evolution. The two different eigenstates with $E^{\pm}$ feature slightly different group velocities which result in that the corresponding wave packets slowly drift away from each other. Coherence leading to oscillatory interference is maintained provided these wave packets have a substantial overlap which requires that $  |v^{+}_g-v^{-}_g|\,L_d \ll \sigma$ where $\sigma$ is the width of the individual wave packets. This is similar to the case of oscillations of active neutrinos and has been analyzed in detail in ref.\cite{kopp} to which the reader is referred   for further discussion. A detailed analysis of possible decoherence effects requires a firm assessment of the energies and energy differences as well as an estimate of the width of the wave packets, which is ultimately determined by characteristic localization length scale of the parent particle and determined by the experimental setup.


\section{Conclusions, possible cosmological implications and further questions:}

Motivated by their astrophysical, cosmological and phenomenological relevance, their important place in   compelling extensions beyond the Standard Model and  recent proposals to search for heavy neutral leptons, we have studied the production, propagation and decay of nearly degenerate heavy sterile neutrinos with common decay channels.

We have implemented   a non-perturbative field theoretical systematic generalization of the Wigner-Weisskopf theory  ubiquitous in the study of neutral meson mixing,   here   extended to  include  both the  production and the decay into the full dynamics for the general case of sterile neutrinos with a common decay channel. Mixing between them is a consequence of a common set of intermediate states which lead to off-diagonal terms in the self-energies. Within the Wigner-Weisskopf description mixing is manifest in off-diagonal terms in the ``effective Hamiltonian'' that describes the time evolution of the \emph{amplitudes} for the sterile neutrino states.

 Our study focuses  on   heavy sterile neutrinos produced by pseudoscalar meson decay as this is one important   avenue  for possible study in current and future neutrino experiments, however the method may be straightforwardly generalized to  alternative production reactions.

 While the dynamical evolution features similarities with the cases of neutral meson mixing, there are noteworthy differences primarily a consequence of including the dynamics of the production and decay in the treatment.

Although the framework is general, we considered the   case of a ``visible'' leptonic common decay channel $\nu_h,\nu_{h'} \rightarrow e^+ e^- \nu_a$ ($a$ is an active neutrino), as an explicit example of experimental relevance and obtained the (nearly degenerate) complex energies. Interference between the ``mass eigenstates'' are manifest in damped oscillations in the $e^+ e^-$ distribution function   akin to the ``quantum beat'' phenomenon in the radiative decay of multilevel atoms.

 In combination with a wave packet description, we obtain the oscillatory contribution to the number of $e^+ e^-$ pairs within a detector of length $\Delta L_d$ placed at a distance $L_d$ from the production region.

 These oscillations in the decay products would be a telltale signature of mixing between heavy neutral leptons.

\textbf{Possible cosmological implications:}
    The decay width of the propagating modes (see eqn.(\ref{Ghee})) suggest that sterile neutrinos in the mass range $M_h \sim 100\,\mathrm{MeV}$ and with $|U_{eh}|^2\lesssim10^{-7}-    10^{-5}$ feature a lifetime ranging from a few seconds to a few minutes depending on the strength of the mixing matrix elements. If these sterile neutrinos are produced from pion decay shortly after the QCD (hadronization) transition (at $\simeq 10\mu\, s~;~T \simeq 150 \,\mathrm{MeV}$) they may decay into $e^+ e^- \nu_a$ several minutes after the freeze out of  active neutrinos. In this case the  active neutrinos from decay are ``injected'' into the cosmic neutrino background with a non-equilibrium distribution function and cannot thermalize after neutrino freeze out. These extra non-thermal neutrinos would not contribute to Big Bang Nucleosynthesis (BBN) as they are produced well after the time scale for BBN, but \emph{may} modify the effective number of relativistic neutrinos,
  $N_{eff}$,  with a non-equilibrium distribution function.


   While the results obtained here yield insights into this possibility, the formulation introduced in this article is not directly applicable to the cosmological case which requires the time evolution of a density matrix instead of an initial single particle state. Furthermore the production of sterile neutrinos must be studied within the quantum kinetics   from pion decay in the thermal medium and freeze out of the distribution function when the pion abundance becomes suppressed as the temperature decreases during the cosmological expansion.
  This study will be reported elsewhere\cite{lello}.


\textbf{Further questions:} In this  study we focussed on understanding the interference effects between
the nearly degenerate sterile neutrinos and their manifestation in the decay products within a general framework.

 We
did not consider specifically either CP violating transitions, or $|\Delta l| = 2$ transitions in the case of Majorana neutrinos. As pointed out in refs.\cite{revs,pilares,pila1,asa1} CP violation may be resonantly enhanced in the case of nearly degenerate heavy sterile neutrinos, furthermore,   lepton violating transitions are suppressed in the case of small (Majorana)  neutrino masses but may be enhanced by heavy sterile neutrinos in intermediate states. Of particular interest would be possible oscillations in $|\Delta l| =2$ transitions.
Furthermore, a complete assessment of the probability of detection requires to consider specific cases for the production reaction as well as the decay interaction vertex, these determine the explicit form of the matrix elements, the coefficients $\alpha^\pm_{1,2}$ in the superposition (\ref{steripm}), the normalization of the Lorentzian distribution in the coefficients $C_{\pm}$ in (\ref{Cis})   and ultimately the overall normalization factor $\mathcal{N}$ in the final expression (\ref{oscinumee}).
All of these aspects merit further study which will be reported elsewhere.

\acknowledgements{The author   acknowledges partial support from NSF-PHY-1202227.}

\appendix
\section{Equivalence with Dyson-resummed propagators:}\label{sec:props}
In the Schroedinger picture the full quantum state is
\bea \label{fulstate}
|\Psi(\vk,t)\rangle_S & = &  C_\Phi(\vec{k},t)\big|\Phi_{\vec{k}}\rangle + \sum_{\alpha;\vq;i=a,h}\,C^{\alpha\, i}_{I}(\vk,\vq;t)\,\big|\nu_{i,\vq};\,L^\alpha_{\vk-\vq} \rangle \nonumber \\ & + &    \sum_{\alpha;\vq;\{X\};\{\vec{p} \}_X} C^{\alpha\,X}_{F}(\vk,\vq,\{\vp\}_X;t)\,\big|L^\alpha_{\vk-\vq}\,; \{X\} \rangle  +\cdots
\eea
where the coefficients in this expression and those of (\ref{intstate}) are related by
\bea C_\Phi(\vec{k},t) & = &  e^{-iE_\Phi t} A_\Phi(\vec{k},t)~;~C^{\alpha\, i}_{I}(\vk,\vq;t) = e^{-iE^i_I t}\,A^{\alpha\, i}_{I}(\vk,\vq;t)\nonumber \\ &&C^{\alpha\,X}_{F}(\vk,\vq,\{\vp\}_X;t) = e^{-iE^X_F t}\,A^{\alpha\,X}_{F}(\vk,\vq,\{\vp\}_X;t)\,. \label{equiCA}\eea The state(\ref{fulstate}) obeys the Schroedinger equation
\be i\frac{d}{dt}|\Psi(\vk,t)\rangle_S = -i(H_0+H_I)|\Psi(\vk,t)\rangle_S \label{Seqn}\ee the equations for the coefficients are obtained by projection in a similar fashion as in section (\ref{sec:form}), with the same notation as in   section (\ref{sec:form}) (see eqns. (\ref{energies}-\ref{Mdj}))  and neglecting the momenta arguments in the coefficients, we obtain
\bea  \dot{C}_{\Phi}(t) && = -i E_\phi \, {C}_{\Phi}(t)-i \sum_{\alpha,\vq,a} {M^{\alpha a}_\mathcal{P}}^* \,C^{\alpha a}_I(t) \nonumber \\&& ~~  -i \sum_{\alpha,\vq,h=1,2} {M^{\alpha\,h}_\mathcal{P}}^* \,C^{\alpha\,h}_I(t)
  ~;~ C_{\Phi}(0)=1  \label{dotCfi1s}\\
 \dot{C}^{\alpha a}_I(t) && = -iE^a_I \,  {C}^{\alpha a}_I(t)-i \,{M^{\alpha a}_\mathcal{P}} \, C_{\Phi}(t)~~;~~{C^{\alpha a}_I}(0)=0 \label{dotCI1a} \\
  \dot{C}^{\alpha h}_I(t) && = -iE^h_I \,  {C}^{\alpha h}_I(t) -i \,{M^{\alpha h}_\mathcal{P}} \, C_{\Phi}(t) -i \sum_{h'=1,2} \delta \mathcal{E}_{hh'}  \,  {C}^{\alpha h'}_I(t) \nonumber \\ &&- i\sum_{\{X\};\{\vp\}_X} {M^{h\,X}_\mathcal{D}}^* \,C^{\alpha\,X}_F(t)  ~;~ {C^{\alpha\,h}_I}(0)=0 ~,~h=1,2 \,,\label{dotCI1s} \\
  \dot{C}^{\alpha X}_F (t) && = -i E^X_F \, {C}^{\alpha X}_F (t) -i \sum_{h=1,2}  {M^{h\,X}_\mathcal{D}}   \,C^{\alpha\,h}_I(t)~~;~~{C^{\alpha\,X} _F}(0)=0 \,.\label{dotCF1s}\eea This hierarchy of coupled differential equations becomes  a set of coupled algebraic equations by Laplace transform, defining
  \be \widetilde{C}(s) = \int_0^\infty e^{-st}\, C(t)\, dt \,,\label{lapla}\ee for all the coefficients, we find beginning from the bottom up
  \be  \widetilde{C}^{\alpha X}_F (s)   =   -i \frac{\Big[{M^{1\,X}_\mathcal{D}}   \,\widetilde{C}^{\alpha\,1}_I(s)+{M^{2\,X}_\mathcal{D}}   \,\widetilde{C}^{\alpha\,2}_I(s)\Big]}{s+i E^X_F} \,,\label{CFXlap} \ee introducing this solution into the Laplace transform of equations (\ref{dotCI1s}), we find
 \be \Bigg[
       \begin{array}{cc}
         s+iE^1_I+i\widetilde{\Sigma}_{11}(s) & i\widetilde{\Sigma}_{12}(s) \\ \\
         i\widetilde{\Sigma}_{21}(s) & s+iE^1_I+i\widetilde{\Sigma}_{22}(s) \\
       \end{array}
     \Bigg]\Bigg(
              \begin{array}{c}
                \widetilde{C}^{\alpha\,1}_I(s) \\ \\
                \widetilde{C}^{\alpha\,2}_I(s) \\
              \end{array}
           \Bigg) = -i \, \widetilde{C}_{\Phi}(s) \, \Bigg(\begin{array}{c}
                                                                                   M^{\alpha 1}_\mathcal{P} \\ \\
                                                                                   M^{\alpha 2}_\mathcal{P}
                                                                                 \end{array}
                                     \Bigg) \label{Cmtx} \ee
where
\be i\widetilde{\Sigma}_{ij}(s)  =    \sum_{\{X\};\{\vp\}_X} \frac{ {M^{iX}_\mathcal{D}}^*\,{M^{jX}_\mathcal{D}}}{s+iE^X_F}+i\, \delta\,\mathcal{E}_{ij}\,. \label{siglap}\ee

The first term in $\widetilde{\Sigma}_{ij}(s)$ corresponds to the intermediate states $\{X\}$, fig. (\ref{fig:selfie}) shows the self-energy for the case of a common three body decay channel.

The solution of the set of equations (\ref{Cmtx}) is given by
\be \Bigg(
              \begin{array}{c}
                \widetilde{C}^{\alpha\,1}_I(s) \\ \\
                \widetilde{C}^{\alpha\,2}_I(s) \\
              \end{array}
           \Bigg) = -i \widetilde{\mathds{G}}(s) \Bigg(\begin{array}{c}
                                                                                   M^{\alpha 1}_\mathcal{P} \\ \\
                                                                                   M^{\alpha 2}_\mathcal{P}
                                                                                 \end{array}
                                     \Bigg) \, \widetilde{C}_{\Phi}(s)\,, \label{Cssol}\ee

           where

\be  \widetilde{\mathds{G}}(s)  =  \frac{1}{D(s)}  \, \Bigg[
       \begin{array}{cc}
         s+iE^1_I+i\widetilde{\Sigma}_{22}(s) & -i\widetilde{\Sigma}_{12}(s) \\ \\
         -i\widetilde{\Sigma}_{21}(s) & s+iE^1_I+i\widetilde{\Sigma}_{11}(s) \\
       \end{array}
     \Bigg] \label{Vdefs} \ee with

\be D(s) = \Bigg[(s+iE^1_I+i\widetilde{\Sigma}_{11}(s))(s+iE^2_I+i\widetilde{\Sigma}_{22}(s))- \widetilde{\Sigma}_{12}(s)\widetilde{\Sigma}_{21}(s) \Bigg]\,. \label{det} \ee

For the amplitudes corresponding to the active neutrinos we find for their Laplace transform
\be   {\widetilde{C}}^{\alpha a}_I(s) = -i \frac{{M^{\alpha a}_\mathcal{P}}}{s+iE^a_I}  \,\widetilde{C}_{\Phi}(s)\,. \label{acampla}\ee Introducing (\ref{Cssol}) and (\ref{acampla}) into the Laplace transform of  (\ref{dotCfi1s}) we find
\be \widetilde{C}_{\Phi}(s) = \frac{1}{s+iE_\Phi+ i\Sigma_\Phi(s)}\label{Cfisol}\ee where
\be \Sigma_\Phi(s) =  \Sigma^{(a)}_\Phi(s)+\Sigma^{(s)}_\Phi(s)  \label{sigmafi}\ee with
\be \Sigma^{(a)}_\Phi(s) = -i\sum_{\alpha,\vq,a} \frac{|{M^{\alpha a}_\mathcal{P}}|^2}{s+iE^a_I} \label{sigmaac} \ee and
\be \Sigma^{(s)}_\Phi(s) = -i\sum_{\alpha,\vq} \Big({M^{\alpha 1}_\mathcal{P}}^*~,~{M^{\alpha 2}_\mathcal{P}}^*  \Big)~\widetilde{\mathds{G}}(s)~ \Bigg(\begin{array}{c}
                                                                                   M^{\alpha 1}_\mathcal{P} \\ \\
                                                                                   M^{\alpha 2}_\mathcal{P}
                                                                                 \end{array}
                                     \Bigg) \label{sigster}\ee are the contributions to the $\Phi$ self-energy from the active $(a)$ and sterile $(s)$ neutrinos, this latter contribution highlights the nature of the resonant heavy neutrino states because $\mathds{G}(s)$ includes the self-energy corrections in the mixed heavy neutrino propagator.

The time evolution is obtained from the anti-Laplace transform, namely for all the amplitudes
\be C(t) = \int_{\mathcal{C}} e^{st}\,\widetilde{C}(s) \,\frac{ds}{2\pi i} \label{antilap}\ee where $\mathcal{C}$ is the Bromwich contour running parallel to the imaginary axis in the complex s plane to the right of all the singularities of $\widetilde{C}(s)$. Decaying states are described by complex poles in $\widetilde{C}(s)$ with a negative real part, therefore along the Bromwich contour $s=i \omega +\epsilon$ with $ -\infty \leq \omega \leq \infty~,~\epsilon \rightarrow 0^+$ and
\be C(t) = \int_{-\infty}^{\infty} e^{i\omega t} \,\widetilde{C}(s=i\omega+\epsilon)\,\frac{d\omega}{2\pi}\,. \label{Cofts}\ee
In perturbation theory $\widetilde{C}_\Phi(s=i\omega + \epsilon)$ features a complex pole near $\omega \sim -E_\Phi$, writing to leading order
\be \Sigma_\Phi(s=-iE_\phi + \epsilon) = \Delta E_\Phi-i \frac{\Gamma_\Phi}{2} \label{sigphipole} \ee it follows that $\widetilde{C}_\Phi(s=i\omega + \epsilon)$  near this pole is of the Breit-Wigner form\footnote{Again we neglect wave function renormalization.}
\be  \widetilde{C}_\Phi(s=i\omega + \epsilon) \simeq -\frac{i}{\omega+E^R_\Phi-i\frac{\Gamma_\Phi}{2}}~~;~~E^R_\Phi = E_\Phi + \Delta E_\Phi \label{BWphi} \ee and
\be C_\Phi(t) = e^{-iE^R_\Phi t}\,e^{-\frac{\Gamma_\Phi}{2}t}\,. \label{cfioft} \ee From the convolution theorem for Laplace transforms we find
\be \Bigg(
              \begin{array}{c}
                 {C}^{\alpha\,1}_I(t) \\ \\
                 {C}^{\alpha\,2}_I(t) \\
              \end{array}
           \Bigg) = -i \int^t_0\,\mathds{G}(t-t') \Bigg(\begin{array}{c}
                                                                                   M^{\alpha 1}_\mathcal{P} \\ \\
                                                                                   M^{\alpha 2}_\mathcal{P}
                                                                                 \end{array}
                                     \Bigg) \,  {C}_{\Phi}(t')\,dt'\,, \label{Cssoloft}\ee where

\be  \mathds{G}(t) =  \int_{-\infty}^{\infty}\widetilde{G}(\omega)\, e^{i\omega t} \,\frac{d\omega}{2\pi}~~;~~\widetilde{G}(\omega)\equiv \widetilde{G}(s=i\omega+\epsilon)\,. \label{Gofts}\ee

To simplify notation we define
\bea && \mathcal{E}_{11}(\omega) \equiv E_1+E_\alpha+\widetilde{\Sigma}_{11}(\omega)~~;~~\mathcal{E}_{22}(\omega) \equiv E_1+E_\alpha+\widetilde{\Sigma}_{22}(\omega) \label{diagsome}\\ &&
\mathcal{E}_{12}(\omega) \equiv \widetilde{\Sigma}_{12}(\omega)~;~\mathcal{E}_{21}(\omega) \equiv \widetilde{\Sigma}_{21}(\omega)\label{offdiagome}\eea with
\be \widetilde{\Sigma}_{ij}(\omega) \equiv \widetilde{\Sigma}_{ij}(s=i\omega+\epsilon) \,.\label{sigis}\ee It follows that the analytic continuation
\be \widetilde{\mathds{G}}(\omega) = - \frac{1}{\big[\omega-\omega^+(\omega)\big]\big[\omega-\omega^-(\omega)\big]}
 \Bigg[
       \begin{array}{cc}
         \omega+\mathcal{E}_{22}(\omega) & - \mathcal{E}_{12}(\omega) \\ \\
         -\mathcal{E}_{21}(\omega) & \omega+\mathcal{E}_{11}(\omega) \\
       \end{array}
     \Bigg] \label{gofomega}\ee where
     \be \omega^\pm(\omega) = -\frac{1}{2} \Bigg\{(\mathcal{E}_{11}(\omega)+\mathcal{E}_{22}(\omega)) \pm \Bigg[(\mathcal{E}_{11}(\omega)-\mathcal{E}_{22}(\omega))^2+4 \mathcal{E}_{12}(\omega)\mathcal{E}_{21}(\omega)\Bigg]^{1/2} \Bigg\} \label{omepm}\ee
     The propagator $\widetilde{G}(\omega)$ features (simple) complex poles at
     \be \omega = \omega^\pm(\omega) \,, \label{polis}\ee these self-consistent conditions can be solved perturbatively. Again there are two cases:

      \vspace{2mm}

      \textbf{a):} $\mathbf{|E_1 -E_2| \gg \widetilde{\Sigma}_{ij}(E_{1,2})}$ for which we find that
     \be \omega^+ = -\mathcal{E}_{11}(E^1_I) = -[\lambda^+ + E_\alpha]~~;~~\omega^- = -\mathcal{E}_{22}(E^2_I) = -[\lambda^- + E_\alpha]\,, \label{casea}\ee where
     $\lambda^\pm$ is given by (\ref{dellar}).

 \vspace{2mm}

      \textbf{b):} $\mathbf{|E_1 -E_2| \lesssim \widetilde{\Sigma}_{ij}(\overline{E})}$ In this case we can set $\omega = \overline{E}+E_\alpha$ in the arguments of the self-energies to leading order $\mathcal{O}(g^2)$ and again we find
\be \omega^+ =  -[\lambda^+ + E_\alpha]~~;~~\omega^-     = -[\lambda^- + E_\alpha]\,, \label{casea2}\ee where in this case
     $\lambda^\pm$ are given by (\ref{lambdascomp}) with $E_{1,2} \rightarrow \overline{E}$ in the arguments of the self-energies.

     In both cases straightforward contour integration finally yields
     \be \mathds{G}(t) = \frac{ e^{-iE_\alpha t}}{\lambda^+ - \lambda^-} \Bigg\{e^{-i\lambda^+ t}\,  \Bigg[
       \begin{array}{cc}
         \lambda^+ -{E}_{22}  &   {E}_{12}  \\ \\
         {E}_{21}  & \omega+\lambda^+ -{E}_{11} \\
       \end{array}
     \Bigg] + e^{-i\lambda^- t}\,  \Bigg[
       \begin{array}{cc}
         \lambda^- -{E}_{22}  &   {E}_{12}  \\ \\
         {E}_{21}  &  \lambda^- -{E}_{11} \\
       \end{array}
     \Bigg]  \Bigg\} \label{goftfini}\ee where the $E_{ij}$ are the same as in (\ref{massmtx}) with $E_{1,2} \rightarrow \overline{E}$ in the self-energies. With the result (\ref{cfioft}) it is now straightforward to find the coeeficients $C^{\alpha 1}_I(t),C^{\alpha 2}_I(t)$ from eqn. (\ref{Cssoloft}). Using the results of appendix (\ref{sec:identities}) we confirm the Wigner-Weisskopf result (\ref{finA}) with (\ref{Vpmsolu}) to leading order, thereby establishing that the Wigner Weisskopf approximation is indeed equivalent to the Dyson resummation of the propagators in terms of the self-energy. This is  a non-perturbative result that generalizes the simpler case analyzed in ref.\cite{cascade} and establishes the relation to the field theoretical propagator approach studied in ref.\cite{pilares}.

\section{Useful identities:}\label{sec:identities}
The eigenvalue equation (\ref{eigen})
\be \Bigg( \begin{array}{cc}
                                      H_{11}  & H_{12}   \\
                                      H_{21}  & H_{22}
                                    \end{array} \Bigg)  \Big(
                       \begin{array}{c}
                         \alpha^{\pm}_1 \\
                         \alpha^{\pm}_2 \\
                       \end{array}
                    \Big) = \lambda^{\pm} \Big(
                       \begin{array}{c}
                         \alpha^{\pm}_1 \\
                         \alpha^{\pm}_2 \\
                       \end{array}
                    \Big) \label{eigen22} \ee we obtain
\be \lambda^{\pm} = \frac{1}{2} \Bigg\{(H_{11}+H_{22})\pm \Bigg[(H_{11}-H_{22})^2+4 H_{12}H_{21} \Bigg]^{1/2}\Bigg\} \label{lambdas} \ee from which it follows that
\be \lambda^- - H_{11} = - (\lambda^+ - H_{22}) \,,\label{eap1}\ee and
\be \frac{\lambda^+-H_{11}}{\lambda^- - H_{11}} = \frac{\lambda^- - H_{22}}{\lambda^+ - H_{22}} \label{eap2}\, , \ee along with the ratios
\be \frac{\alpha^-_1}{\alpha^-_2} = \frac{\lambda^- - H_{22}}{H_{21}}~~;~~  \frac{\alpha^+_2}{\alpha^+_1} = \frac{\lambda^+ - H_{11}}{H_{12}} \,.\label{ratios} \ee

With these results, after straightforward algebra we find the following identities:
\be \alpha^+_1\, \widetilde{M}^{\alpha +}_\mathcal{P} = \frac{(\lambda^+ - H_{22}) M^{\alpha 1}_\mathcal{P}  + H_{12}  M^{\alpha 2}_\mathcal{P} }{\lambda^+ - \lambda^-} \label{ide1}\ee

\be \alpha^-_1\, \widetilde{M}^{\alpha -}_\mathcal{P} = -\frac{(\lambda^- - H_{22}) M^{\alpha 1}_\mathcal{P}  + H_{12}  M^{\alpha 2}_\mathcal{P} }{\lambda^+ - \lambda^-} \label{ide2}\ee

\be \alpha^+_2\, \widetilde{M}^{\alpha +}_\mathcal{P} = \frac{(\lambda^+ - H_{11}) M^{\alpha 2}_\mathcal{P}  + H_{21}  M^{\alpha 1}_\mathcal{P} }{\lambda^+ - \lambda^-} \label{ide3}\ee

\be \alpha^-_2\, \widetilde{M}^{\alpha -}_\mathcal{P} = -\frac{(\lambda^- - H_{11}) M^{\alpha 2}_\mathcal{P}  + H_{21}  M^{\alpha 1}_\mathcal{P} }{\lambda^+ - \lambda^-} \label{ide4}\ee
 The important aspect is that these products are independent of the normalization of $\alpha^\pm_{1,2}$.

\section{Derivation of eqn. (\ref{intfinifor}).}\label{app:deriv}

$A^F_X(t)$ given by (\ref{Afinat}) can be written in obvious notation as $A^+(t)+A^-(t)$ corresponding to the first and second lines in (\ref{Afinat}). The interference terms are
\be (A^+(t))^* A^-(t) + \mathrm{c.c.} \label{produc} \ee

To simplify notation we introduce the auxiliary quantities
\be \tau^{\pm} = \Big(\alpha^\pm_1 \widetilde{M}^{\alpha \pm}_\mathcal{P}\, M^{1X}_\mathcal{D}+ \alpha^\pm_2 \widetilde{M}^{\alpha \pm}_\mathcal{P}\, M^{2X}_\mathcal{D}\Big) \ee and
\be \mathcal{E} = \overline{\mathcal{E}}+E_\alpha - E_\Phi~~; ~~ \eta = E^X_F - E_\Phi ~~;~~ \Delta^\pm_\gamma = \Gamma_\Phi - \Gamma^\pm \ee where $\overline{\mathcal{E}},\delta$  have been defined in eqn. (\ref{defini2}). The interference term $(A^+(t))^* A^-(t)$ is  given by
\be  (A^+(t))^* A^-(t) = \frac{\tau^+ \tau^-}{\big(\mathcal{E}+\delta-i\frac{\Delta^{+}_\gamma}{2} \big) \big(\mathcal{E}-\delta+i\frac{\Delta^{-}_\gamma}{2} \big)}\,\Bigg\{ (1)+(2)+(3)+(4)\Bigg\}   \ee where
\be(1) = \frac{\Big(e^{-i\eta t}\,e^{-\frac{\Gamma_\Phi}{2}t}-1 \Big)}{\Big(\eta-i\frac{\Gamma_\Phi}{2} \Big)}\,\frac{\Big(e^{ i\eta t}\,e^{-\frac{\Gamma_\Phi}{2}t}-1 \Big)}{\Big(\eta+i\frac{\Gamma_\Phi}{2} \Big)} \,,\label{uno}  \ee
\be (2) = \frac{\Big(e^{-i(\eta-\mathcal{E})t}\,e^{i\delta t}\,e^{-\frac{\Gamma^+}{2} t} -1 \Big)}{\Big(\eta -\mathcal{E}-\delta -i \frac{\Gamma^+}{2} \Big)}\,\frac{\Big(e^{ i(\eta-\mathcal{E})t}\,e^{i\delta t}\,e^{-\frac{\Gamma^-}{2} t}-1 \Big)}{\Big(\eta -\mathcal{E}+\delta +i \frac{\Gamma^-}{2} \Big)}   \label{dos} \ee
\be (3) = -\frac{\Big(e^{-i\eta t}\,e^{-\frac{\Gamma_\Phi}{2}t}-1 \Big)}{\Big(\eta-i\frac{\Gamma_\Phi}{2} \Big)}\,\frac{\Big(e^{ i(\eta-\mathcal{E})t}\,e^{i\delta t}\,e^{-\frac{\Gamma^-}{2} t}-1 \Big)}{\Big(\eta -\mathcal{E}+\delta +i \frac{\Gamma^-}{2} \Big)} \label{tres}\ee
\be (4) = -  \frac{\Big(e^{-i(\eta-\mathcal{E})t}\,e^{i\delta t}\,e^{-\frac{\Gamma^+}{2} t} -1 \Big)}{\Big(\eta -\mathcal{E}-\delta -i \frac{\Gamma^+}{2} \Big)}\,\frac{\Big(e^{ i\eta t}\,e^{-\frac{\Gamma_\Phi}{2}t}-1 \Big)}{\Big(\eta+i\frac{\Gamma_\Phi}{2} \Big)} \label{cuatro} \ee

Out of these four contributions, it is only contribution $(2)$ that survives at long time with an oscillatory behavior on long time scales, $(1)$ does not feature oscillatory interference and $(3),(4)$ feature rapidly varying phases $e^{\pm i  \mathcal{E}t}$ but not interference terms and decay on   time scales $1/\Gamma_\Phi$.

The denominators in $(2)$ feature resonances at $E^{\pm} \simeq E^X$   precisely when the heavy sterile neutrinos (the correct eigenstates) can decay into the common channel, in the narrow width limit these resonant denominators become energy conserving delta functions.  We can obtain the coefficient functions of these delta functions by integrating in the complex $\eta$ plane and extracting the residues at the complex poles. In the nearly degenerate limit $\overline{\mathcal{E}} \gg \delta$ and understood as a distribution that is integrated over a density of states that is insensitive to the energy difference $\delta$ we find
\be (2) = \frac{2\pi}{\overline{\Gamma}}\,\frac{\Big[1-e^{2i\delta t}\,e^{-\overline{\Gamma}t} \Big]}{\Big[1-i \frac{2\delta}{\overline{\Gamma}} \Big]}\,\delta(\eta - \mathcal{E})\,.\label{2fini}\ee

Similarly, in the narrow width limit and in the nearly degenerate case the product
\be \frac{1}{\big(\mathcal{E}+\delta-i\frac{\Delta^{+}_\gamma}{2} \big) \big(\mathcal{E}-\delta+i\frac{\Delta^{-}_\gamma}{2} \big)} \propto \delta(\mathcal{E})\,,\ee to find the proportionality factor we integrate in the complex $\mathcal{E}$ plane extracting the residues at the complex poles and find this product (as a distribution integrated over smooth density of states that is insensitive to the energy difference $\delta$) with the result
\be  \frac{1}{\big(\mathcal{E}+\delta-i\frac{\Delta^{+}_\gamma}{2} \big) \big(\mathcal{E}-\delta+i\frac{\Delta^{-}_\gamma}{2} \big)} = \frac{2\pi \,\delta(\mathcal{E})}{\Gamma_\Phi+\overline{\Gamma}+2i\delta}\,. \label{dist}\ee

This result can be easily understood as follows: in the narrow width limit
\be \frac{1}{\big(\mathcal{E}+\delta-i\frac{\Delta^{+}_\gamma}{2} \big) \big(\mathcal{E}-\delta+i\frac{\Delta^{-}_\gamma}{2} \big)} = \frac{1}{-2\delta+i(\Gamma_\Phi+\overline{\Gamma})}\,\Bigg[i\pi \Big((\delta(\mathcal{E}+\delta) +(\delta(\mathcal{E}-\delta)\Big)+ \mathcal{P}\Big(\frac{1}{(\mathcal{E}+\delta) }-\frac{1}{(\mathcal{E}-\delta) } \Big) \Bigg] \label{difa}\ee upon integrating over a density of final states that is insensitive to the energy difference $\delta$ the result (\ref{dist}) above follows. The same analysis applies to the result (\ref{2fini}).

 Therefore the final result for the interference term is
\be  (A^+(t))^* A^-(t) =  {\tau^+ \tau^-} \frac{2\pi \, \delta(\overline{\mathcal{E}}+E_\alpha-E_\Phi)}{\Big[\Gamma_\Phi+\overline{\Gamma}+2i\delta\Big]} \frac{2\pi\,\delta(\overline{\mathcal{E}}-E^X)}{\Big[\overline{\Gamma} -2i \delta \Big]}\,\Big[1-e^{2i\delta t}\,e^{-\overline{\Gamma}t} \Big] \ee

\end{document}